\documentclass[iop]{emulateapj}
\def\axpu0142{4U~0142+61}
\def\rxsj1708{1RXS~J170849.0-400910}
\def\axpe1048{1E~1048.1-5937}
\def\be1547{1E~1547.0-5408}
\def\ce1841{1E~1841-045}
\def\de2259{1E~2259+586}
\def\acxouj1647{CXOU~J164710.2-455216}
\def\bcxouj1714{CXOU~J171405.7-381031}
\def\xtej1810{XTE~J1810-197}
\def\asgr0418{SGR~0418+5729}
\def\bsgr0501{SGR~0501+456}
\def\csgr1900{SGR~1900+14}
\def\swiftj1822{{\it Swift}~J1822.3-1606}

\begin{document}

\shorttitle{STEMS3D}
\title{X-ray Perspective of the Twisted Magnetospheres of Magnetars}


\author{Shan-Shan Weng$^{1}$, Ersin G\"o\u{g}\"u\c{s}$^{1}$, Tolga G\"uver$^{2}$, \& Lin Lin$^{3}$}

\affil{$^{1}$ Sabanc\i~University, Faculty of Engineering and Natural
  Sciences, Orhanl\i ~Tuzla 34956 Istanbul Turkey}

\affil{$^{2}$ Istanbul University, Science Faculty, Department of Astronomy and
Space Sciences, Beyaz{\i}t, 34119, Istanbul, Turkey}

\affil{$^{3}$ Fran\c{c}ois Arago Centre, APC, Universit\'{e} Paris Diderot,
CNRS/IN2P3 13 rue Watt, 75013 Paris, France}

\begin{abstract}

Anomalous X-ray pulsars (AXPs) and soft gamma-ray repeaters (SGRs) are
recognized as the most promising magnetar candidates, as indicated by their
energetic bursts and rapid spin-downs. It is expected that the strong magnetic
field leaves distinctive imprints on the emergent radiation both by affecting
the radiative processes in atmospheres of magnetars and by scattering in the
upper magnetospheres. We construct a self-consistent physical model that
incorporates emission from the magnetar surface and its reprocessing in the
three-dimensional (3D) twisted magnetosphere using a Monte Carlo technique. The
synthetic spectra are characterized by four parameters: surface temperature kT,
surface magnetic field strength $B$, magnetospheric twist angle $\Delta\phi$,
and the normalized electron velocity $\beta$. We also create a tabular model
(STEMS3D) and apply it to a large sample of {\it XMM-Newton} spectra of
magnetars. The model successfully fits nearly all spectra, and the obtained
magnetic field for 7 out of the 11 sources are consistent with the values
inferred from the spin-down rates. We conclude that the continuum-fitting by
our model is a robust method to measure the magnetic field strength and
magnetospheric configuration of AXPs and SGRs. Investigating the multiple
observations of variable sources, we also study the mechanism of their spectral
evolution. Our results suggest that the magnetospheres in these sources are
highly twisted ($\Delta\phi > 1$), and the behavior of magnetospheric twisting
and untwisting is revealed in the 2002 outburst of \de2259.

\end{abstract}


\section{Introduction}
Anomalous X-ray pulsars (AXPs) and soft gamma-ray repeaters (SGRs) form an
intriguing class of isolated neutron stars (NSs). They share the similar
observational properties: (1) relatively slow spin periods ($P \sim 2-12$ s);
(2) large spin-down rates ($\dot{P} \sim 10^{-14}-10^{-10}$ s s$^{-1}$); (3)
high persistent X-ray luminosity ($L_{\rm X} \approx 10^{34}-10^{36}$ erg
s$^{-1}$), which significantly exceeds their rotation power for most sources;
and (4) different types of X-ray variability; i.e., long-term X-ray flux
enhancements and short energetic bursts (see \citealt{mereghetti08}, and
\citealt{rea11} for recent reviews). These behaviors can be understood within
the magnetar scenario -- extremely strong magnetic fields of some NSs giving
rise to the observed exotic manifestations \citep{thompson95,thompson96}. There
are now 28 magnetars or magnetar candidates known
\footnote{\protect\url{http://www.physics.mcgill.ca/$\sim$pulsar/magnetar/main.html}}
\citep{olausen14, zhou14}.

Although recent multi-wavelength observations of magnetar activities
significantly widened our views, the most essential knowledge on magnetars is
provided by the emission properties of their X-ray radiation. The X-ray spectra
of SGRs and AXPs below 10 keV are soft and can be empirically described by a
power-law (PL) model with photon index $\Gamma \sim 2-4$ and a thermal
component with temperature $kT \sim 0.5$ keV \citep{mereghetti05,
mereghetti06}. Also note that it has been revealed that some magnetars exhibit
a very hard spectrum ($\Gamma \sim 1$) above 10 keV \citep[see,
e.g.,][]{kuiper04, kuiper06, den08a, den08b}. \cite{beloborodov13} suggested
that the hard X-ray emissions attribute to surface emissions scattered by the
relativistic outflows around magnetars, which also make a non-negligible
contribution below 10 keV. Studying the spectra of persistent emission,
\cite{marsden01} reported that the hardness of the spectra increases with the
spin-down rate $\dot{\nu}$. With more accumulated data, \cite{kaspi10} showed
that the spectral indices in quiescence are not only correlated with
$\dot{\nu}$ but also the magnetic field $B$ inferred from timing analyses.
Magnetars occasionally emit short ($\sim 0.1$ s) bursts, during which the
plausible line features have been also detected in some sources
\citep[e.g.,][]{strohmayer00}. If such features are due to the proton cyclotron
absorption, the inferred magnetic field somehow deviates from the value
estimated from the spin-down rate \citep[e.g.,][]{woods05, tiengo13, an14}. An
independent method is thus required to evaluate the magnetic field of
magnetars.

In the magnetar framework, the X-ray radiation can be physically interpreted as
the NS surface emission scattered by the plasma in the magnetosphere
\citep[see, e.g.,][]{thompson02, lyutikov06, fernandez07, tong10}.
\cite{thompson02} suggested that the internal magnetic field of magnetars is
highly twisted and anchored in the highly conducting crust of an NS. In
starquakes, the strong internal magnetic field deforms the crust, and as a
result the footpoints of the external magnetic field lines are displaced, i.e.,
the magnetic field outside the star -- magnetosphere is also twisted.
Alternatively, a twisted magnetosphere can also be aroused by continued
footpoint motions \citep{beloborodov09}. The twisted external magnetic field
can support electric currents that are much larger than the Goldreich-Julian
current \citep{thompson02, lyutikov06}. The charge carriers provide large
optical depth to resonant cyclotron scattering (RCS) so that soft photons can
be Compton up-scattered and form a PL high-energy tail. Both one-dimensional
(1D) and three-dimensional (3D) RCS models had been developed \citep[see
e.g.,][]{lyutikov06, fernandez07}, and offered promising fits for the X-ray
spectra of SGRs/AXPs \citep{rea08, zane09}. Because the non-resonant scattering
contributes a negligible fraction, both 1D and 3D RCS models only take the
resonant scattering into consideration. For $\sim$ keV photons, the resonant
scattering occurs in the layer where the magnetic field decays to $\sim
10^{11}$ G regardless of the NS surface magnetic field. As a consequence, the
essential information of SGRs/AXPs -- the strength of the magnetic field---is
not incorporated in both 1D and 3D RCS models since the scattering process is
insensitive to the magnetic field below its cyclotron frequency ($\geq$ 100 keV
for a magnetar magnetic field).

Besides the magnetospheric scattering, the surface emissions are also distorted
in strongly magnetized ($B \geq 10^{13}$ G) NS atmospheres due to the effect of
vacuum polarization and proton cyclotron resonances, by affecting the
interactions between the photons and the plasma \citep[e.g.,][]{ho03, ozel03}.
Thus, the spectral profile of surface emission strongly depends on the magnetic
field. Taking into account the combined effects of the magnetar atmosphere and
its magnetosphere, \cite{guver07} developed the 1D surface thermal emission and
magnetospheric scattering (STEMS) model, which can successfully fit the soft
X-ray spectra of both steady and variable magnetars \citep{guver07, guver12,
ng11, lin12}. Additionally, the STEMS model uncovers the information concering
the magnetic field. However, because the STEMS model treats the scattering
region as a plane-parallel slab, it cannot offer information about the geometry
of the magnetosphere.

The long-term monitoring observations revealed that the X-ray emissions of most
magnetars are variable \citep[e.g.,][]{rea14}, in the extreme cases, namely the
transient magnetars (e.g., \xtej1810, and \be1547), the overall flux can be
enhanced up to three orders of magnitude brighter than their quiescent level
\citep[e.g.,][]{ibrahim04, mereghetti09, kaneko10}. Studying X-ray behaviors in
different flux levels would help us to explore the nature of magnetar, i.e.,
the evolution of magnetospheres \citep{beloborodov09} and the magneto-thermal
evolution of NSs \citep{pons12, vigano13}. Based on {\it XMM-Newton}
observations, \cite{zhu08} found that the afterglow of the \de2259's 2002
outburst followed by a PL decay and the hardness is strongly correlated with
the flux. Alternatively, the flux relaxation of \xtej1810's 2003 outburst is
best described by an exponential decay and the surface temperature became cold
in the meantime \citep{gotthelf05}. Diverse observational phenomena (see
\citealt{rea11} and reference therein) suggest that X-ray activities could be
driven by different mechanisms, either the twisted magnetospheres
\citep{thompson02} or the deep crustal heating \citep{lyubarsky02}.

In this work, we carry out 3D Monte Carlo simulations of the emitted photons
from the surface of a highly magnetized NS propagation in a twisted
magnetosphere. The physical bases of our model and the Monte Carlo method are
described in Section 2. The model properties are presented in Section 3. We
also numerically calculate model spectra and create a tabular model, which is
further applied to the X-ray spectra of magnetars (Section 4). The ultimate
goal of this paper is to better understand the surface and magnetospheric
properties of magnetars and their evolutions using our model. In Section 5, we
discuss the spectral modeling results and their implications.

\section{Models}

In this section, we first outline the physical processes in a strongly
magnetized atmosphere, and then scatter the surface emission in the 3D twisted
magnetosphere. The gravitational redshift effect is also considered in our
model.

\subsection{Surface emission of magnetars}
In strong magnetic fields ($B \geq 10^{13}$ G), the vacuum polarization, a
quantum electrodynamics phenomenon, can affect radiative transfer in magnetized
plasmas by modifying polarization modes and the opacities of the normal modes
\citep[e.g.,][]{zane01, ho03, ozel03}. The proton cyclotron resonance is
another phenomenon that affects the propagation of photons in a magnetized
plasma. Following the methods discussed in \cite{ozel01, ozel03}, the radiative
equilibrium models are constructed for the fully ionized hydrogen plasmas,
taking into account the effects of vacuum polarization and ion cyclotron lines.
We use a modified Feautrier method for the solution of the angle- and
polarization-mode dependent radiative transfer problem and ensure radiative
equilibrium with a temperature correction scheme based on the Lucy--Uns\"{o}ld
algorithm. The surface emission spectrum can be defined by just two parameters:
the effective temperature of atmosphere kT and the surface magnetic field
strength $B$.

\subsection{Resonant cyclotron scattering in 3D twisted magnetosphere}

The Monte Carlo techniques are quite suitable to handle photon scattering in
complicated 3D configurations and the 3D RCS Monte Carlo code has already been
built by \cite{fernandez07} and \cite{nobili08}. Both of these works considered
axisymmetric globally twisted magnetospheres and seed photons as being the
canonical blackbody (BB) emissions. Nevertheless, these codes allow for an
arbitrary distribution of seed photons, velocity distribution of the charged
particles, and magnetic field geometry.

We build our Monte Carlo code following an approach that is similar to the one
discussed in \cite{nobili08}. The fundamental difference is the description of
seed photons. (1) The original BB emissions are distorted after travelling
through strongly magnetized atmospheres, mostly carried by extraordinary mode
photons \citep{ozel03}. In our work, this surface emission is adopted as the
seed photons, which are homogeneously injected from the stellar surface with a
wavevector pointing in the radial direction. Note that under the influence of
an ultrastrong magnetic field, the surface emissions are expected to be
anisotropic and their temperature varies with latitude
\citep[e.g.,][]{heyl98,bernardini11}. However, the study on the anisotropic
surface emissions is beyond the scope of this paper and will be reported
elsewhere.  (2) The globally twisted magnetosphere has a self-similar and
force-free construction ($\mathbf{j} \times \mathbf{B} = 0$), which is uniquely
characterized by the net twist angle of the field lines anchored close to the
two magnetic poles, $\Delta\phi$. (3) The current density $\mathbf{j}$ is
deduced from the Ampere's law ($\triangledown \times \mathbf{B} =
(4\pi/c)\mathbf{j}$) under the force-free assumption. Following the work in
\cite{nobili08}, we consider the effects of bulk velocity and also thermal
velocity distribution of charge carriers, assuming a 1D Maxwellian distribution
at a given temperature superimposed to a bulk motion. In order to reduce
parameters in the model, the electron temperature $kT_{\rm e}$ is derived from
the bulk velocity (and further halved) by assuming the equipartition between
thermal and bulk kinetic energy. As a result, the number density of electrons
($n_{\rm e}$) is a function of the twist angle and the average electron
velocity. (4) We neglect both the non-resonant scattering and electron recoil
effects, which are not important in the soft X-ray band. The spectrum is
calculated in the non-relativistic regime with the simplified resonant
cross-section \citep[Equation (10) in][]{nobili08}, which is independent of
magnetic field and frequency.

\begin{figure}
\begin{center}
\includegraphics[scale=0.4]{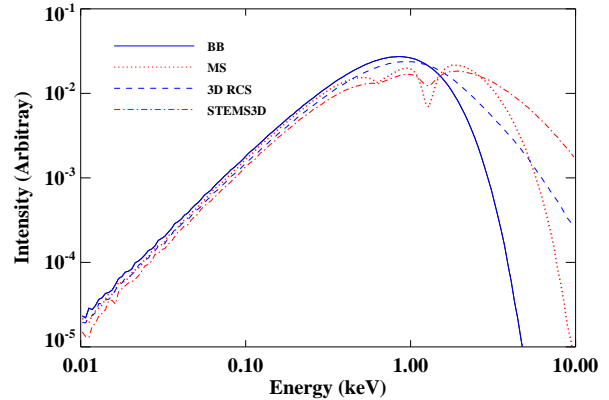} \caption{The BB (kT $= 0.3$ keV, blue
solid line) and the magnetar surface (MS, $kT = 0.3$ keV and $B = 10^{14}$ G,
red dotted line) emissions are scattered in a 3D twisted magnetosphere
($\Delta\phi = 1.0$ and $\beta=0.3$), resulting in the spectra of the 3D RCS
(blue dashed line) and the STEMS3D (red dotted-dashed line), respectively.
\label{model}}
\end{center}
\end{figure}

As photons propagate in the 3D twisted magnetosphere, we check whether they can
freely escape out of the magnetosphere in every step \citep{fernandez07,
nobili08}. If the photon frequency together with moving direction still allow
resonant scattering, as soon as the integrated scattering depth $\tau_{\rm s}$
satisfies $\tau_{\rm s} \geq -lnU$, where $U$ is a random number in the range
[0,1], a scattering is triggered, and the photon frequency, direction, and
polarization are updated. Then, the photon continues to travel in the
magnetosphere until another scattering occurs or it escapes to infinity
(observers). We also refer to \cite{nobili08} for more details on the
scattering process. The spectra are calculated by evolving 5,000,000 photons
and characterized by four parameters: surface temperatures $kT$, magnetic field
strength at the poles $B$, magnetospheric twist angles $\Delta\phi$, and the
normalized electron velocity $\beta = v/c$, where $c$ is the speed of light. An
example spectrum (kT $= 0.3$ keV, $B = 10^{14}$ G, $\Delta\phi = 1.0$, and
$\beta=0.3$) is shown in Figure \ref{model}.

\subsection{Gravitational redshift effect}
\begin{figure}
\begin{center}
\includegraphics[scale=0.4]{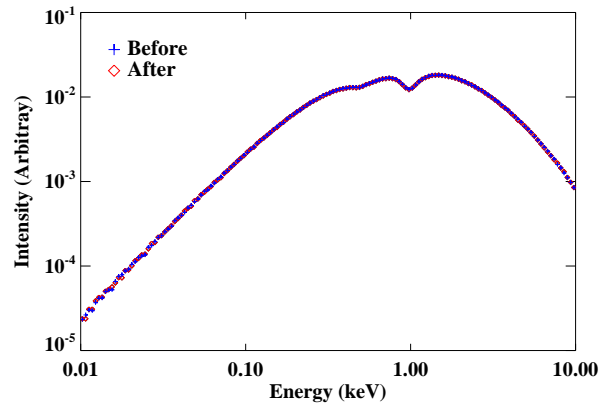} \caption{Example
spectrum (kT $= 0.3$ keV, $B = 10^{14}$ G, $\Delta\phi = 1.0$, and $\beta =
0.3$) with the gravitational redshift correction ($z = 0.306$) performed before
(blue plus) and after (red diamond) the magnetospheric scattering.
\label{redshift}}
\end{center}
\end{figure}

Because of the strong gravitational field of NSs, the gravitational redshift
effect is important and should be corrected for the photons close to NSs, but
is negligible for the photons that are far away from NSs ($>$ several times of
the NSs radii). According to General Relativity, the photons experience a
redshift in gravitational fields by a factor of $1+z =
\frac{1}{\sqrt{1-\frac{2GM}{Rc^{2}}}}$ assuming Schwarzschild geometry. In
principle, the gravitational redshift should be corrected for the surface
emission, before the magnetospheric scattering. Since the scattering optical
depth is independent of frequency in a self-similar magnetosphere
\citep{thompson02, fernandez07}, it allows us to correct the spectra after the
magnetospheric scattering. In order to test this hypothesis, we calculated the
spectra for kT $= 0.3$ keV, $B = 10^{14}$ G, $\Delta\phi = 1.0$, and $\beta =
0.3$ with the gravitational redshift correction of $z = 0.306$ (corresponding
to an NS with mass 1.4 $M_{\odot}$ and $R_{\rm NS} = 10$ km) performed before
(red line in Figure \ref{redshift}) and after (blue line) the magnetospheric
scattering, and find that both spectra are identical. Therefore, it becomes
feasible and convenient to do the correction in {\it XSPEC} (after scattering,
\S 4).

\section{Model properties}

The emerging radiation from the surface of a highly magnetized NS is expected
to be modified by both vacuum polarization and proton cyclotron resonance. At
magnetar-type field strengths, the resulting spectrum is spectrally harder than
a Planckian and would exhibit proton line features in the soft-X-ray band
(Figure \ref{model}). The emerging photons further gain energy by multiple
scattering in the twisted magnetosphere, manifesting themselves as the high
energy tail. To illustrate this point, we performed the scattering of a BB ($kT
=0.3$ keV) as well as the magnetar surface emission ($kT =0.3$ keV and $B =
10^{14}$ G) in a 3D twisted magnetosphere ($\Delta\phi =1.0$ and $\beta=0.3$).
We present these spectra in Figure \ref{model}. We also calculated the spectrum
of the 3D RCS in the same way as the spectrum of STEMS3D model, but with a
single temperature BB adopted for the seed photons.

\begin{figure}
\begin{center}
\includegraphics[scale=0.4]{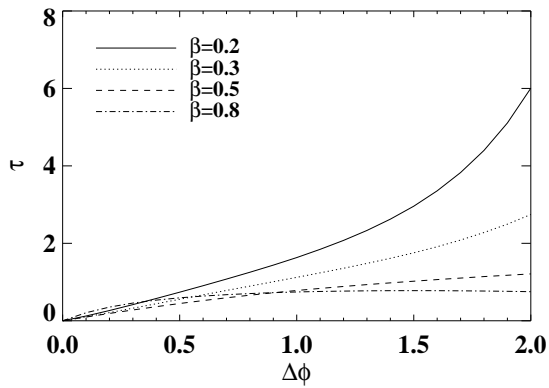} \caption{Angle-averaged optical
depth ($\tau$) vs. $\Delta\phi$ and different values of $\beta$: 0.2 (solid),
0.3 (dotted), 0.5 (dashed), and 0.8 (dashed-dotted).\label{optical}}
\end{center}
\end{figure}

\begin{figure*}
\begin{center}
\includegraphics[scale=0.4]{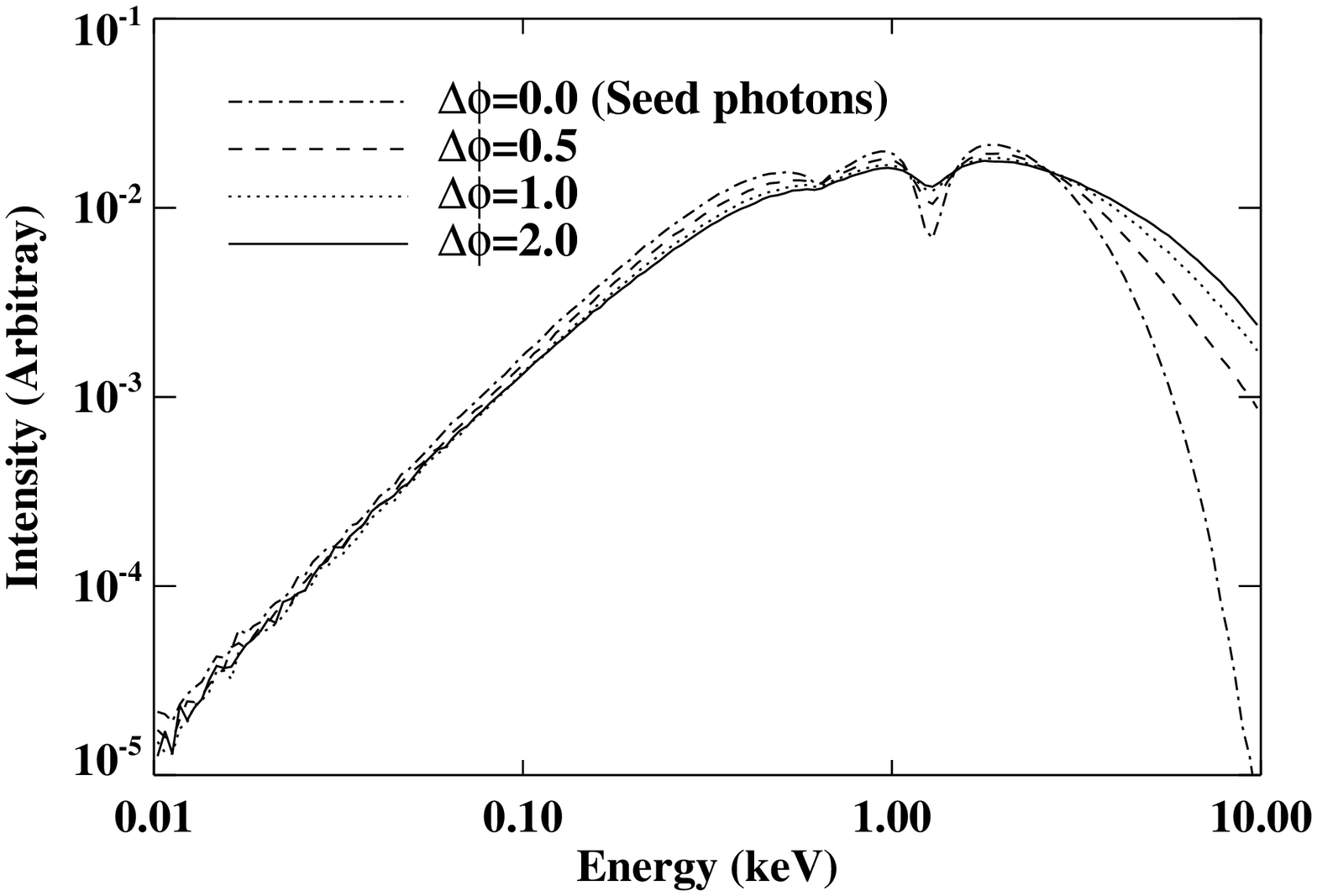}
\includegraphics[scale=0.4]{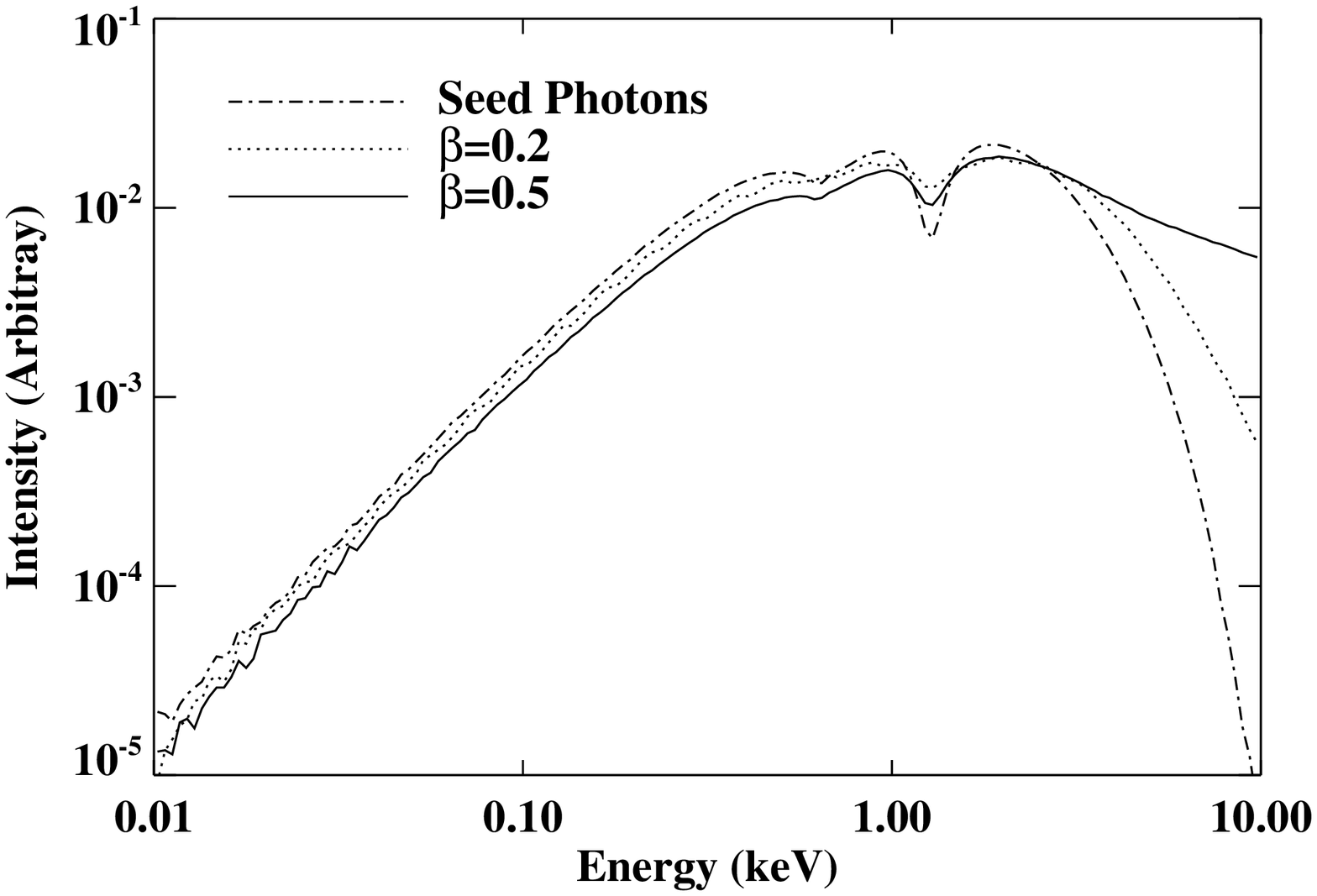} \caption{Left: model
spectra (without gravitational redshift correction) for $B = 10^{14}$ G, kT $=
0.3$ keV, $\beta = 0.3$ and different values of $\Delta\phi$. Right: computed
spectra for $B = 10^{14}$ G, kT $= 0.3$ keV, $\Delta\phi = 1.0$ and different
values of $\beta$. \label{spectra}}
\end{center}
\end{figure*}

Unlike the STEMS model, in which the optical depth is independent of the
particle velocity \citep{lyutikov06, guver07}, the optical depth in the STEMS3D
is determined by both the degree of magnetospheric twist and the velocity of
charged particles. In Figure \ref{optical}, we show that the variation of
angle-averaged mean number of scattering per photon ($\tau$) with respect to
$\Delta\phi$ and $\beta$. Figure \ref{spectra} illustrates the effects on the
spectral shape of varying twist angle and electron velocity. It is clearly seen
that the level of upscattering of emergent thermal photons increases with
$\Delta\phi$ and in particular with $\beta$.

\cite{lyutikov06} pointed out that the deep line features present in the
surface spectra can be smoothed out with a large scattering optical depth.
However, investigating the spectra obtained with a narrow velocity distribution
($\bar{\beta}=0.5$, $\Delta\beta = 0.1$) and large twist ($\Delta\phi = 1.0$),
\cite{fernandez07} argued that the line strength can be reduced by a factor of
$\sim 2$; however, a significant line feature remained as a result of an
optical depth in the 3D twisted magnetosphere that was not very large. As
already mentioned, the optical depth in 3D twisted magnetosphere depends not
only on $\Delta\phi$ but increases sharply with decreasing $\beta$ (Figure
\ref{optical}). As we show in Figure \ref{spectra}, we find that the STEMS3D
model can indeed wash out the line features if the twist angle is large
($\Delta\phi \geq 1$) and the electron speed is slow ($\beta \leq 0.3$) enough.

\section{Application to soft X-ray spectra of magnetars}

We generated model X-ray spectra in the 0.01--10.0 keV range using the
numerical model described in \S 3. We used the following parameter space for
our numerical grid: the surface temperature $kT = 0.1-0.6$ keV (step 0.1 keV),
surface magnetic field strength $B =10^{14}-10^{15}$ G (step $10^{14}$ G),
magnetospheric twist angle $\Delta\phi = 0.0-2.0$ radian (step 0.1 radian), and
the normalized electron velocity $\beta = 0.1-0.9$ (step 0.1). We then employed
the routine {\it wftbmd},
\footnote{\protect\url{http://heasarc.gsfc.nasa.gov/docs/heasarc/ofwg/docs/general/ \\
  modelfiles\_memo/modelfiles\_memo.html.}},
to create a tabular model (named STEMS3D.mod) which can be implemented into the
standard package for X-ray spectral analysis, {\it XSPEC}. This tabular model
is further applied to the observational data of magnetars.

\subsection{Observations and data analysis}

The main objective of our efforts here is to better understand the surface and
magnetospheric properties of magnetars using the STEMS3D model. For this
investigation, we selected the sources that were bright enough ($f_{\rm 0.5-10
keV} \geq 10^{-12}$ erg cm$^{-2}$ s$^{-1}$) to allow statistically significant
spectral results and were located in relatively low interstellar absorption
($nH \le 5\times10^{22}$ cm$^{-2}$) regions of our Galaxy. When $nH >
5\times10^{22}$ cm$^{-2}$, X-ray emissions below 2 keV are strongly absorbed,
making the parameters of the STEMS3D model unconstrained, even for the bright
source SGR 1806-20. The flux of the serendipitously discovered magnetar, 3XMM
J185246.6+003317, during its active phase is a few times of $10^{-12}$ erg
cm$^{-2}$ s$^{-1}$. However, it was only detected with EPIC-MOS \citep{zhou14},
which has a smaller effective area. We do not include these spectra owing to
the low signal-to-noise ratio (S/N).

Our sample includes bright persistent magnetars that exhibited only subtle
long-term flux variations even when they experienced glitches or emitted
energetic X-ray bursts \axpu0142 \citep{gavriil11}; \rxsj1708 \citep{ssm13};
\ce1841 \citep{ssm14}; and \csgr1900 \citep{israel08}. On the other hand, two
bright persistent sources: \de2259 and \axpe1048 showed long-term X-ray flux
enhancements in connection with X-ray bursts and glitches \citep{kaspi03,
dib09}. Over the last decade, transient magnetars have emerged as a sub-group.
In quiescence, these sources emit X-rays at levels near or below our detection
capabilities. Their X-ray fluxes are enhanced by a factor of 100 or more at the
onset of their outburst episodes \citep{rea14}. We selected \be1547
\citep{israel10}, \swiftj1822 \citep{scholz14}, \bsgr0501 \citep{gogus10,
lin12}, and \xtej1810 \citep{guver07} to investigate the spectral behavior of
transient magnetars. Note that the quiescent spectra of transient magnetars are
excluded even when their fluxes exceed the critical flux level. We discuss the
implications of these spectra and limitations of our model for sources in
quiescence in \S 5. We also include in our sample the longest {\it XMM-Newton}
observation of \bcxouj1714 in 2010. The observational details of the X-ray data
used in this paper are listed in Table \ref{log}. Note that the two short
observations of \ce1841 in 2002 October were performed just two days apart. To
improve counting statistics, we combine these two short observations to one
spectrum.

The data collected with the {\it XMM-Newton} EPIC-pn instrument are reduced
using the Science Analysis System software (SAS) version 12.0.1, and filtered
with the standard criteria: cleaning for background areas, setting FLAG=0 and
PATTERN$\leq$4. We use the SAS task {\it epatplot} to evaluate the pile-up
fraction. For observations showing severe problems with pile-up (Table
\ref{log}), spectra are extracted within annulus regions with the inner radius
$\sim$ 10-20$''$ and ARF files are calculated to correct the missing part of
the point-spread function, thus the correct flux level could still be measured
through spectral fitting. The spectral response files are created using the SAS
tasks {\it rmfgen} and {\it arfgen}.  All spectra are rebinned with the task
{\it specgroup} to have at least 20 counts per bin to enable the use of
chi-square statistics and not to oversample the instrument energy resolution by
more than a factor of three. A 2\% systematic error is added to the data to
account for uncertainties in instrumental calibrations. The spectral analysis
is performed in the 0.5--7.0 keV energy range for the spectra having a low
absorption, while in the 1.0--7.0 keV for those being highly absorbed. We note
that the model-independent residuals are detected in the spectra of \be1547 and
\bcxouj1714 below 1.2 keV; therefore, only the spectra in the energy 1.2--7.0
keV are used (Table \ref{log}). All spectra are fitted with {\it XSPEC} 12.8.1
\citep{arnaud96}. Because the interstellar hydrogen column density (nH) is not
expected to vary during different observations, for each source with more than
one pointing, we fit all available spectra simultaneously and link the nH to
have a common value, and the more updated solar abundances by \cite{lodders03}
are adopted in our work.

\begin{deluxetable*}{clccccc}
\tabletypesize{\tiny} \tablewidth{0pt} \tablecaption{Log of the {\it
XMM-Newton} observations used in this work}  \tablehead{\colhead{Source}&
\colhead{Obs No.} & \colhead{ObsID} & \colhead{Obs Date} & \colhead{Mode} &
\colhead{Net Exposure} &
\colhead{Energy band} \\
\colhead{} & \colhead{} & \colhead{} & \colhead{} &\colhead{} &
\colhead{(ksec)} & \colhead{(keV)}} \startdata \hline
\axpu0142$^{\sharp}$      & Obs1& 0206670101 & 2004 Mar 01 & Timing & 37 & 0.5-7.0 \\
\hline
\rxsj1708$^{\sharp}$     & Obs1& 0148690101 & 2003 Aug 28 & SW & 31 & 0.5-7.0 \\
\hline
\ce1841$^{\sharp}$       & Obs1&  0013340101 & 2002 Oct 05 & LW & 2 & 1.0-7.0 \\
     ...      & ... &  0013340201 & 2002 Oct 07 & LW & 4 & ... \\
\hline
\csgr1900$^{\sharp}$    & Obs1& 0305580101& 2005 Sep 20 & FF & 21 & 1.0-7.0\\
     ...     & Obs2& 0305580201& 2005 Sep 22 & FF & 20 & ...\\
     ...     & Obs3& 0410580101& 2006 Apr 01 & FF & 14 & ...\\
     ...     & Obs4& 0506430101& 2008 Apr 08 & FF & 23 & ...\\
\hline
\axpe1048       & Obs1& 0112780401 &  2000 Dec 28 &  FF & 4  & 0.5-7.0 \\
     ...      & Obs2& 0147860101$^{\dagger}$ &  2003 Jun 16 &  FF & 46 & ...\\
     ...      & Obs3& 0164570301 &  2004 Jul 08 &  Timing & 31 & ...\\
     ...      & Obs4& 0307410201 &  2005 Jun 16 &  SW & 19 & ...\\
     ...      & Obs5& 0307410301 &  2005 Jun 28 &  SW & 16 & ...\\
     ...      & Obs6& 0510010601 &  2007 Jun 14 &  SW & 34 & ...\\
     ...      & Obs7& 0654870101 &  2011 Aug 06 &  FF & 76 & ...\\
     ...      & Obs8& 0723330101$^{\dagger}$ &  2013 Jul 22 &  FF & 54 & ...\\
\hline
\de2259       & Obs1& 0057540101$^{\dagger}$ & 2002 Jan 22 & FF  & 10 & 0.5-7.0 \\
     ...      & Obs2& 0038140101 & 2002 Jun 11 & SW  & 34  & ...\\
     ...      & Obs3& 0155350301 & 2002 Jun 21 & SW  & 18  & ...\\
     ...      & Obs4& 0057540201$^{\dagger}$ & 2002 Jul 09 & FF  & 7  & ...  \\
     ...      & Obs5& 0057540301$^{\dagger}$ & 2002 Jul 09 & FF  & 11 & ... \\
     ...      & Obs6& 0203550301$^{\dagger}$ & 2004 Feb 20 & FF  & 4  & ...\\
     ...      & Obs7& 0203550601 & 2004 Jun 06 & SW  & 5 & ... \\
     ...      & Obs8& 0203550401 & 2004 Jun 22 & SW  & 4 & ... \\
     ...      & Obs9& 0203550501 & 2004 Dec 19 & SW  & 4 & ... \\
     ...      & Obs10& 0203550701 & 2005 Jul 28 & SW  & 3 & ... \\
\hline
\bcxouj1714   & Obs1& 0606020101 & 2010 Mar 17 & FF & 81 & 1.2-7.0 \\
\hline
\xtej1810     & Obs1& 0161360301 &  2003 Sep 08 &  SW & 7  & 0.5-7.0\\
     ...      & Obs2& 0161360501 &  2004 Mar 11 &  LW & 10 & ...\\
     ...      & Obs3& 0164560601 &  2004 Sep 18 &  LW & 23 & ...\\
     ...      & Obs4& 0301270501 &  2005 Mar 18 &  LW & 36 & ...\\
\hline
\bsgr0501     & Obs1& 0560191501 &  2008 Aug 23 &  SW & 34 & 0.5-7.0 \\
     ...      & Obs2& 0552971101 &  2008 Aug 29 &  SW & 17 & ...\\
     ...      & Obs3& 0552971201 &  2008 Aug 31 &  SW & 7 & ...\\
     ...      & Obs4& 0552971301 &  2008 Sep 02 &  SW & 14 & ...\\
     ...      & Obs5& 0552971401$^{\dagger}$ &  2008 Sep 30 &  LW & 28 & ...\\
\hline
\swiftj1822   & Obs1& 0672281801$^{\dagger}$ &  2011 Sep 23 &  LW & 10 & 0.5-7.0 \\
     ...      & Obs2& 0672282701$^{\dagger}$ &  2011 Oct 12 &  LW & 17 & ...\\
     ...      & Obs3& 0672282901 &  2012 Apr 05 &  LW & 23 & ...\\
     ...      & Obs4& 0672283001 &  2012 Sep 08 &  LW & 20 & ...\\
\hline
\be1547$^{\sharp}$       & Obs1&  0560181101$^{\dagger}$ & 2009 Feb 03 & FF & 49 & 1.2-7.0 \\
\hline
\asgr0418     & Obs1&  0610000601 & 2009 Aug 12 & SW & 41 & 0.5-7.0 \\
\hline
\acxouj1647   & Obs1& 0404340101 & 2006 Sep 22 & FF & 40 & 1.0-7.0 \\
     ...      & Obs2& 0410580601 & 2007 Feb 17 & LW & 16 & ...\\
     ...      & Obs3& 0505290201 & 2007 Aug 19 & LW & 22 & ...  \\
     ...      & Obs4& 0505290301 & 2008 Feb 15 & LW & 19 & ... \\
     ...      & Obs5& 0555350101 & 2008 Aug 20 & LW & 28 & ...\\
     ...      & Obs6& 0679380501 & 2011 Sep 27 & LW & 15 & ... \\

\enddata

\tablecomments{Mode: operating mode of the EPIC-pn including full-frame (FF),
large-window (LW), small-window (SW), and timing modes. Net exposure: clean
exposure (in unit of kiloseconds) after background flares excluded. energy
band: Energy band in which spectra are fitted with {\it XSPEC}. \\
$^{\sharp}$: Sources' spectra are fitted by the absorbed (STEMS3D$+$PL) model
with the PL index $\Gamma$ fixed (see the text for more details), and the
others are fitted by a single absorbed STEMS3D model.\\
$\dagger$: Observations suffer problems with pile-up, and spectra are extracted
from annular regions centered on the source position. \\
\label{log}}
\end{deluxetable*}

\begin{figure*}
\begin{center}
\includegraphics[scale=0.7]{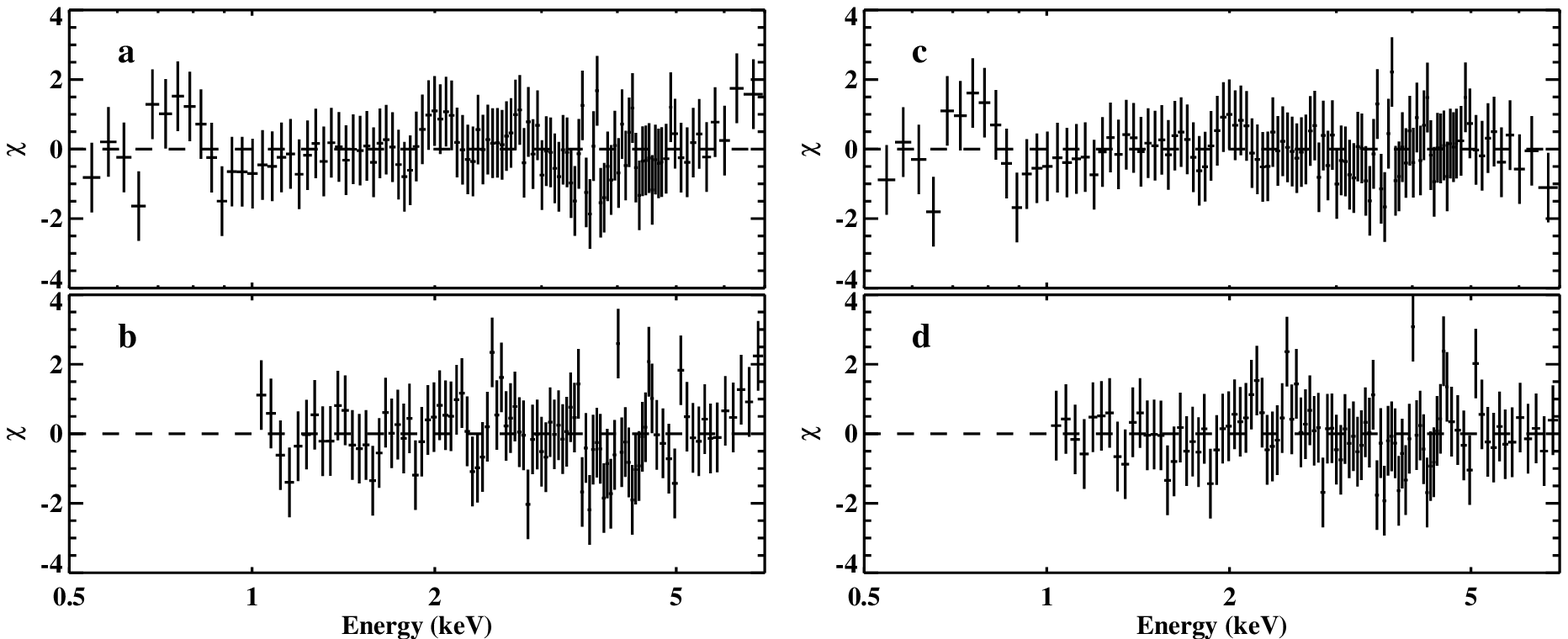} \caption{Spectra of \axpu0142
and \rxsj1708 are fitted with the single absorbed STEMS3D model, and the fit
residuals are shown in panels (a), (b), respectively. Panels (c) and (d) plot
the residuals when the same spectra are fitted with the absorbed (STEMS3D$+$PL)
model. \label{chi}}
\end{center}
\end{figure*}

We first attempt to fit all spectra with an absorbed STEMS3D model assuming the
gravitational redshift parameter of 0.306. We find that spectra of \axpu0142
and \rxsj1708 clearly require an additional hard-X-ray component; therefore,
adding a PL component significantly improved the fit since the mechanism
responsible for the hard-X-ray emission likely contributes a non-negligible
fraction of soft-X-ray emission and naturally affects our fitting (see Figure
\ref{chi}). We fixed the photon index $\Gamma$ to the values obtained from the
fitting to the contemporaneous hard-X-ray data. As the spectra from \axpu0142,
\rxsj1708, \ce1841, and \csgr1900 above 10 keV are relatively stable and can be
fitted with a PL model, we adopt a $\Gamma$ of 0.93 \citep{den08a}, 1.13
\citep{den08b}, 1.32 \citep{kuiper06}, and 1.43 \citep{enoto10a}, respectively,
in addition to STEMS3D in our spectral fitting. Because the hard-X-ray
component in \be1547 exhibited dramatic transient behavior \citep{kuiper12}, we
only analyze the {\it XMM-Newton} observation executed on 2009 February 3, and
we used a PL index $\Gamma$ of 1.41 according to the Suzaku HXD-PIN data on
2009 January 28 \citep{enoto10b}.

\subsection{Spectral analysis, spectral evolution, and correlations}

\subsubsection{Persistent Magnetars}

\begin{figure*}
\begin{center}
\includegraphics[scale=0.5]{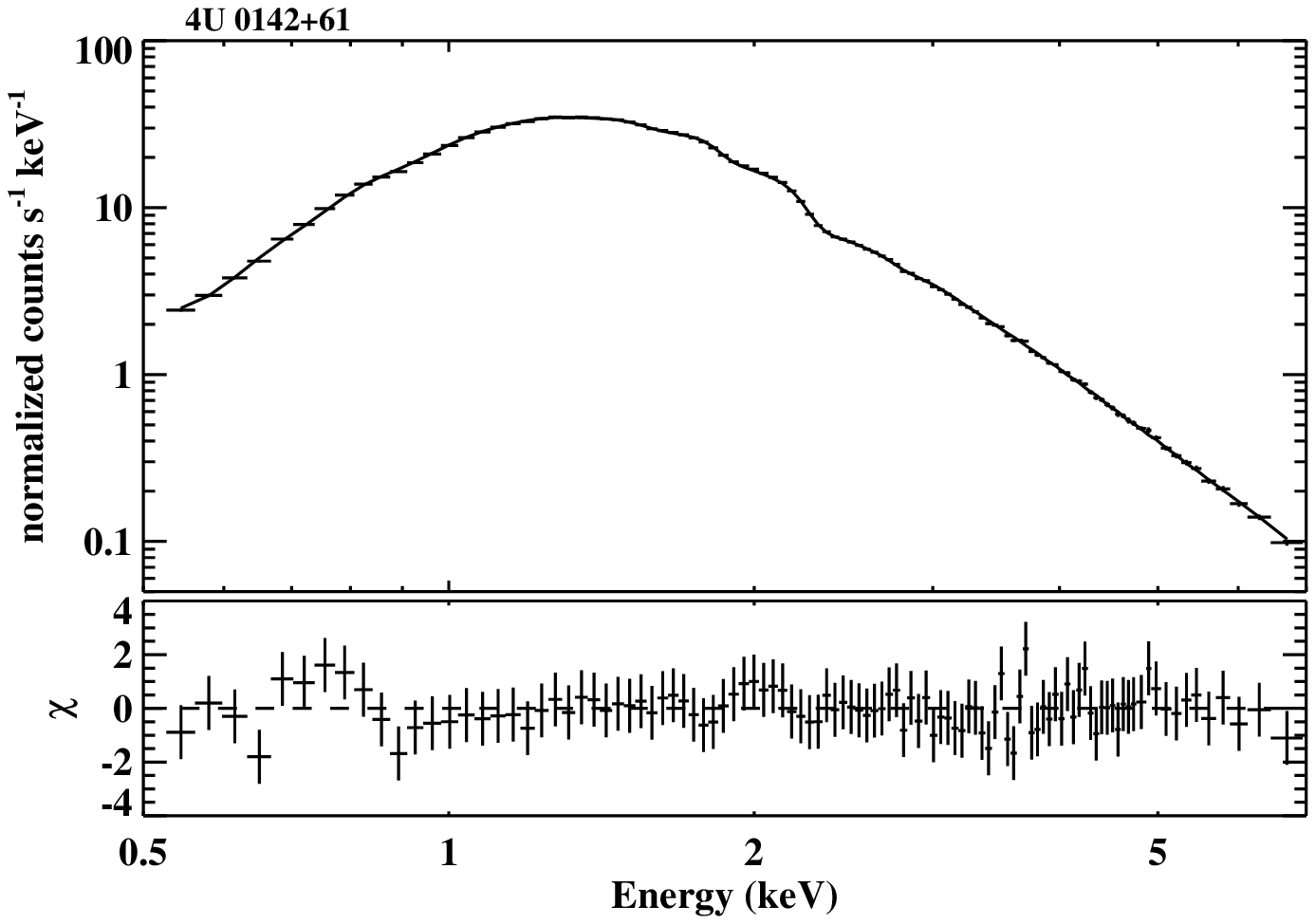}
\includegraphics[scale=0.5]{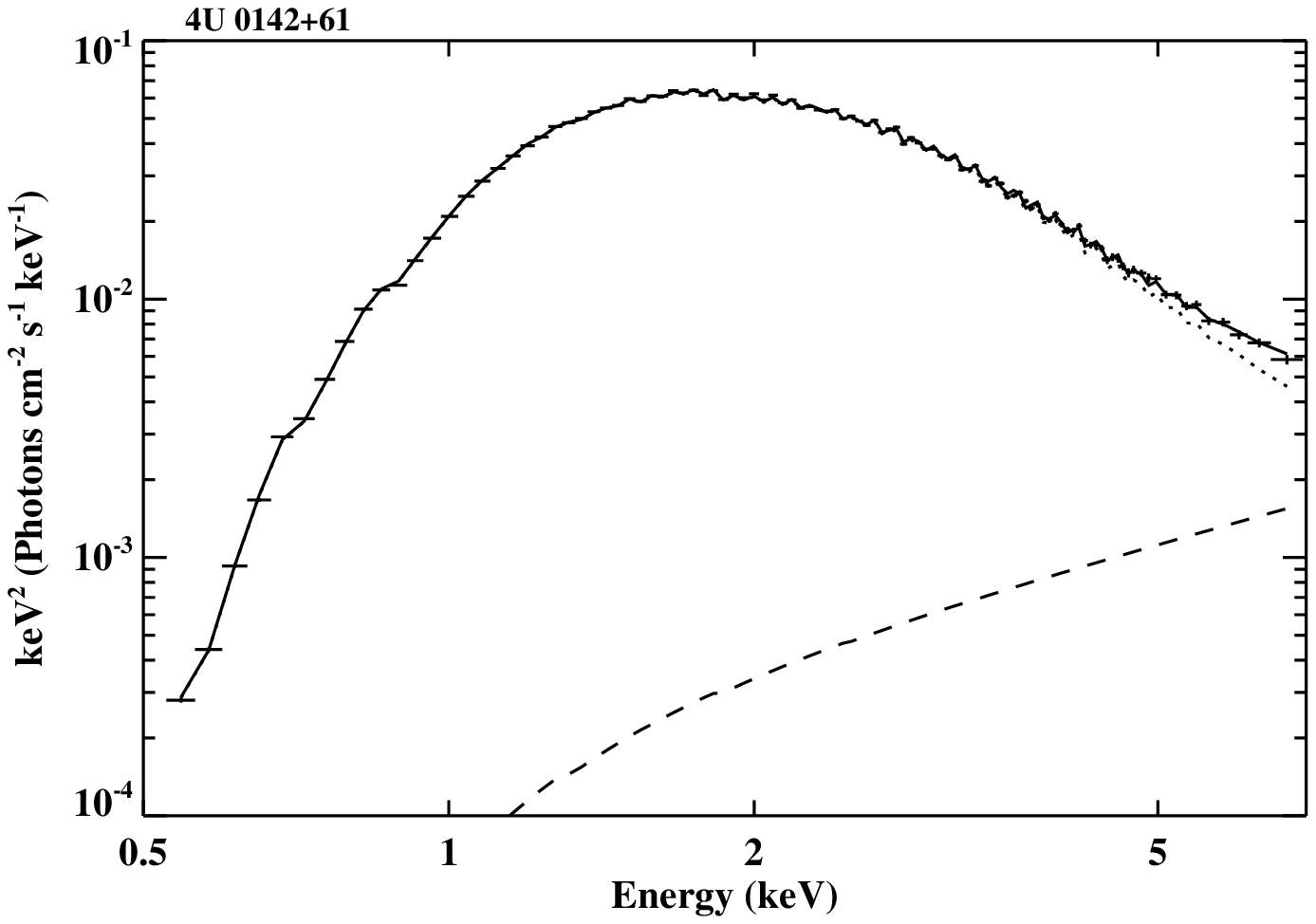}
\includegraphics[scale=0.5]{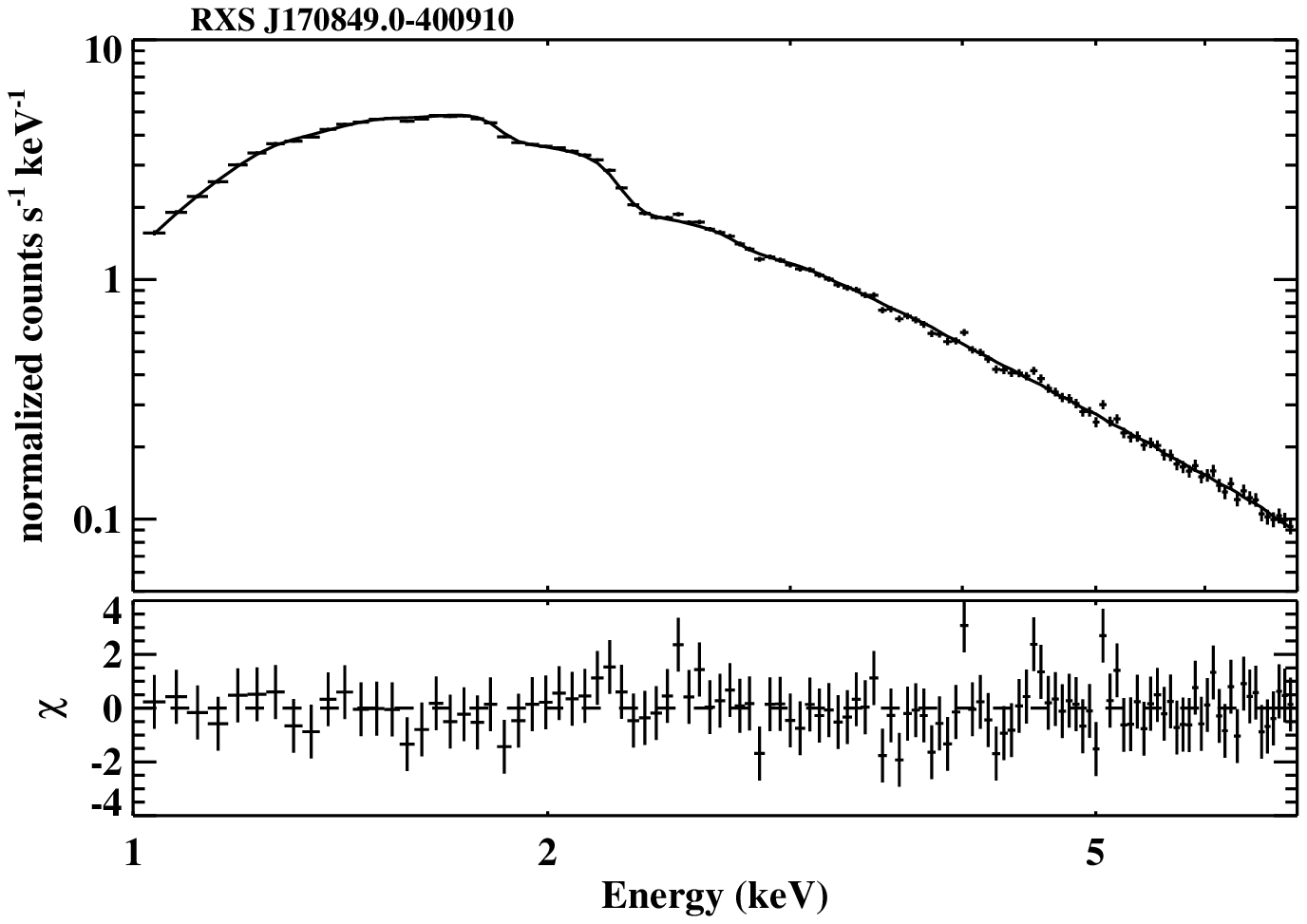}
\includegraphics[scale=0.5]{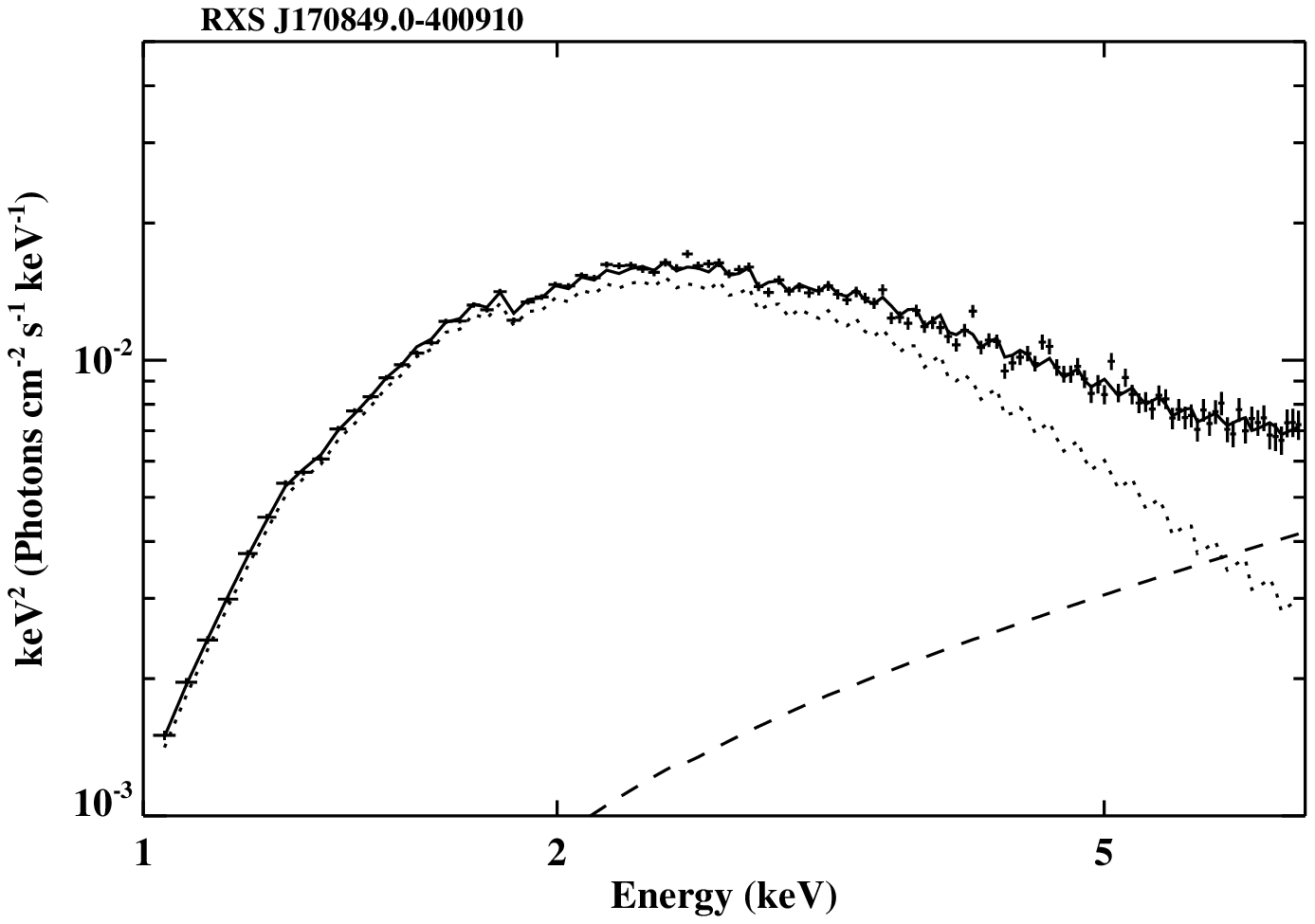}
\includegraphics[scale=0.5]{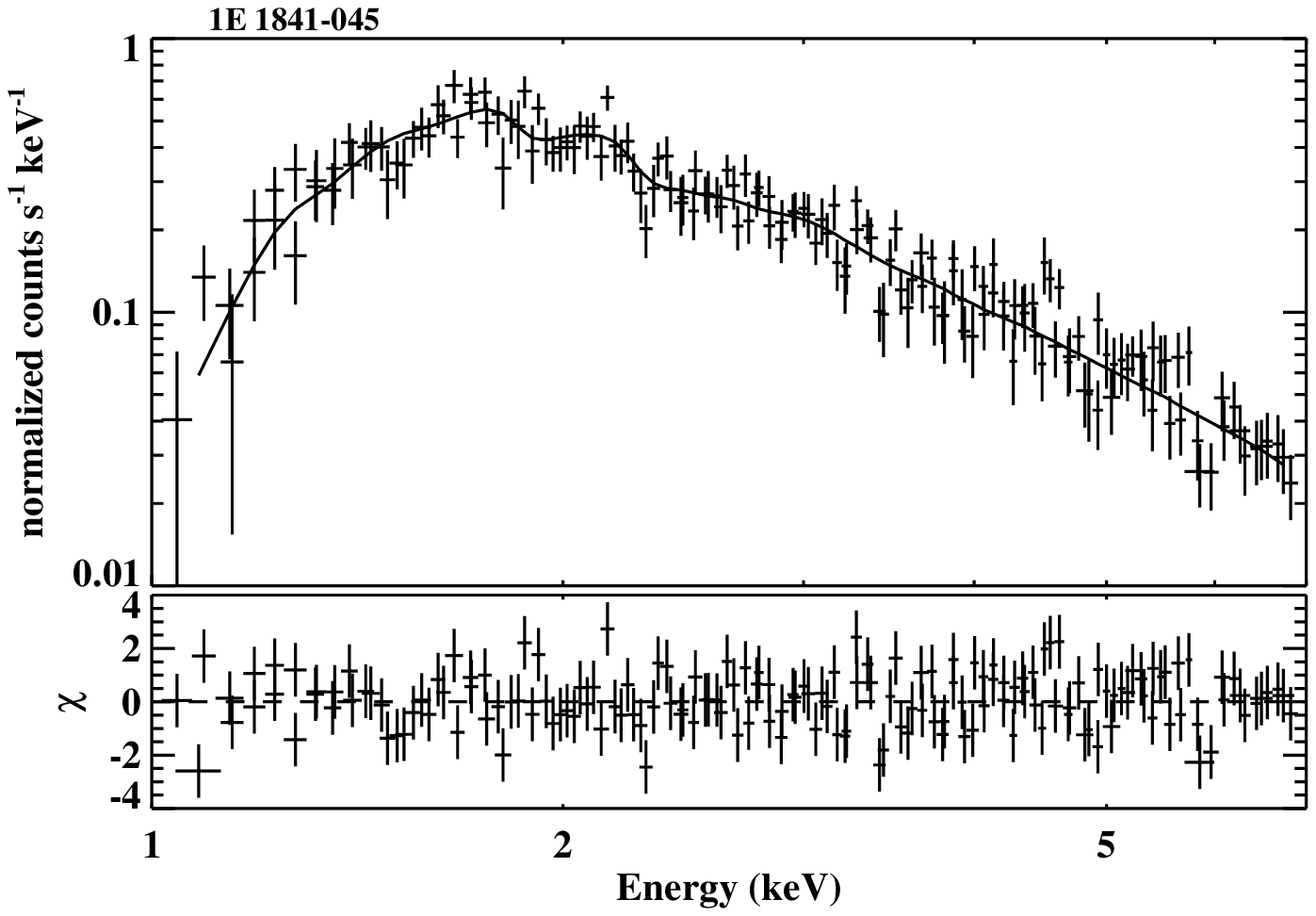}
\includegraphics[scale=0.5]{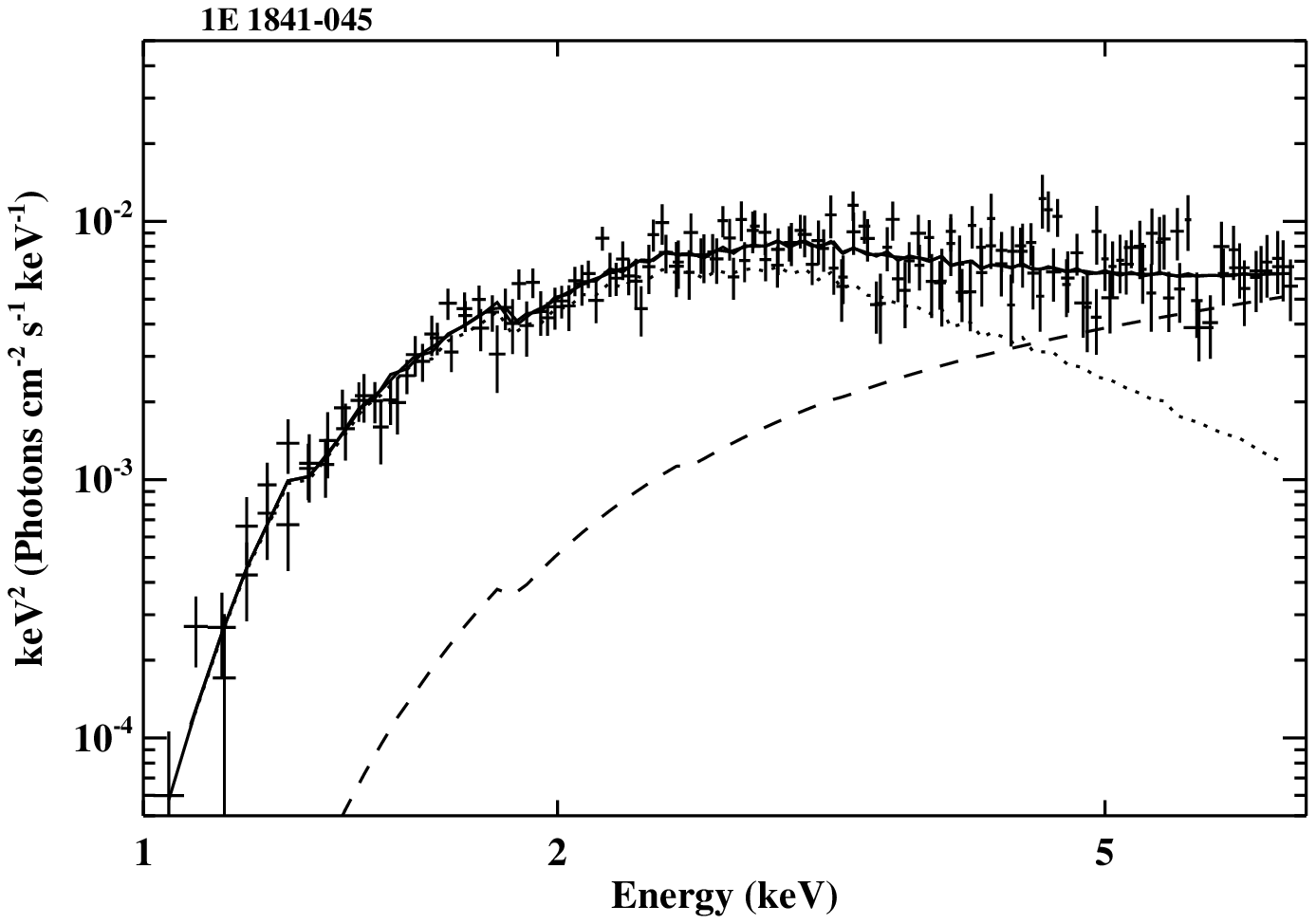}
\caption{Left panels: the X-ray count spectra of \axpu0142, \rxsj1708, and
\ce1841 are fitted by the STEMS3D model (Table \ref{fits_1}) and the residuals
are shown below. Right panels: the corresponding unfolded spectra. The dotted,
dashed, and solid lines in the right panels mark the STEMS3D, PL components,
and the sum, respectively. \label{spec_1}}
\end{center}
\end{figure*}

\axpu0142 is the brightest magnetar and has been one of the most stable X-ray
emitters in X-rays \citep{gonzalez10, wang14}. Its longest {\it XMM-Newton}
spectra in the 0.5$-$7.0 keV band can be modeled perfectly with an absorbed
STEMS3D plus a PL. We find the surface temperature, $kT = 0.33$ keV, and
surface magnetic field, $B = 5.87 \times$10$^{14}$ G (see Table \ref{fits_1}
for all other fit details and Figure \ref{spec_1} for the unfolded spectrum and
best-fit model curve). Note that these surface parameters are consistent with
those that were obtained using STEMS \citep{guver07}. We find that the
magnetospheric electrons are non-relativistic ($\beta$=0.21), and that the
magnetospheric twist angle, $\Delta\phi$ = 1.77, that is close to upper end of
our parameter space. The STEMS3D component contributes more than 98\% of the
total flux below 7 keV.

Since the first glitch activity detected in 1999 \citep{kaspi00}, several
glitches were unveiled in \rxsj1708 \citep{ssm13}. The source emits persistent
radiation in both soft- and hard-X-rays. Fitting the 0.5$-$7.0 keV spectrum, we
find that the surface temperature is relatively high ($kT = 0.49$ keV) and the
other parameters are quite close to those in \axpu0142, i.e., $B = 6.03
\times$10$^{14}$ G, $\Delta\phi$ = 1.82, and $\beta$ = 0.20 (Table
\ref{fits_1}). The STEMS3D component contributes about 88\% of the total
soft-X-ray flux (Figure \ref{spec_1}).

The AXP \ce1841 is located at the center of the X-ray and radio supernova
remnant (SNR) Kes 73 \citep{vasisht97} and has the slowest spin period of
$\sim$ 11.8 s \citep{dib08}. The broadband (1--200 keV) X-ray spectrum can be
described by a BB plus a single PL model \citep{rea08}. As can be seen in
Figure \ref{spec_1}, the PL component only supplies $\sim 8\%$ of the total
flux below 7 keV; however, it starts to overcome the STEMS3D component around 5
keV. The fitting results indicate that the magnetosphere is modestly twisted
($\Delta\phi$ = 0.31) and the electrons have the fastest velocity ($\beta =
0.28$) among all sources studied here, i.e., the smallest optical depth in the
magnetosphere (Figure \ref{optical}).  The model parameters,  however, have
relative large errors because of the short effective exposure ($\sim 6$ ks) in
addition to high interstellar absorption.

\begin{deluxetable*}{cccccccc}
\tabletypesize{\tiny} \tablewidth{0pt} \tablecaption{Spectral fit results of
\axpu0142, \rxsj1708, and \ce1841 } \tablehead{\colhead{Source} & \colhead{nH}
& \colhead{kT} & \colhead{$B$} & \colhead{$\Delta\phi$} &
\colhead{$\beta$} & \colhead{Flux} & \colhead{$\chi^2$/dof} \\
\colhead{}  & \colhead{($10^{22}$ cm$^{-2}$)} & \colhead{(keV)} &
\colhead{($10^{14}$ G)} & \colhead{(rad)} & \colhead{} & \colhead{}
&\colhead{}} \startdata

\hline
\axpu0142  &  $0.73_{-0.05}^{+0.03}$ & $0.33_{-0.03}^{+0.03}$ & $5.87_{-0.49}^{+0.68}$ & $1.77_{-0.06}^{+0.09}$ & $0.21_{-0.01}^{+0.01}$ & $11.7$ & 80.2/122\\
\hline
\rxsj1708  &  $1.50_{-0.04}^{+0.05}$ & $0.49_{-0.02}^{+0.01}$ & $6.03_{-0.14}^{+0.30}$ & $1.82_{-0.08}^{+0.08}$ & $0.20_{-0.01}^{+0.01}$ & $3.26$ & 85.2/109\\
\hline
\ce1841    &  $4.79_{-1.19}^{+0.81}$ & $0.24_{-0.07}^{+0.20}$ & $4.16_{-0.36}^{+0.81}$ & $0.31_{-0.11}^{+0.80}$ & $0.28_{-0.06}^{+0.04}$ & $1.61$ & 178.4/161\\
\enddata

\tablecomments{Flux: 0.5--7.0 keV absorbed flux in units of $10^{-11}$ erg
cm$^{-2}$ s$^{-1}$. All errors are in the 90\% confidence level.
\label{fits_1}}
\end{deluxetable*}

\bcxouj1714 is associated with the SNR CTB 37B and probably with a TeV source
HESS~J1713-381 \citep{aharonian08}. \bcxouj1714 was identified as the youngest
magnetar with a characteristic age of $\sim 950$ yr \citep{sato10}. There have
been only modest flux variations (less than a factor of two) reported with two
Chandra and one {\it XMM-Newton} observations during 2007 February and 2010
March \citep{sato10}. Currently, no model provides a good fit to its X-ray
spectra below $\sim$1.2 keV, likely due to the contamination of diffuse
nonthermal X-rays around the source (in the SNR). We, therefore, investigate
its 1.2--7.0 keV spectrum with a single STEMS3D model, and obtain nH $=
4.39_{-0.23}^{+0.33}$ $\times 10^{22}$ cm$^{-2}$, $kT = 0.51_{-0.02}^{+0.04}$
keV, $B = 9.73_{-0.59} \times$10$^{14}$ G, $\Delta\phi = 0.80_{-0.37}^{+0.14}$,
$\beta = 0.25_{-0.03}^{+0.05}$, absorbed flux in 0.5--7.0 keV of $1.5$ $\times
10^{-12}$ erg cm$^{-2}$ s$^{-1}$, and $\chi^{2}/dof = 121.4/98$. Owing to the
lowest flux and a large value of nH, we cannot obtain the upper error of $B$,
which turns out to be the strongest surface magnetic field estimate among all
sources.

\begin{figure*}
\begin{center}
\includegraphics[scale=0.5]{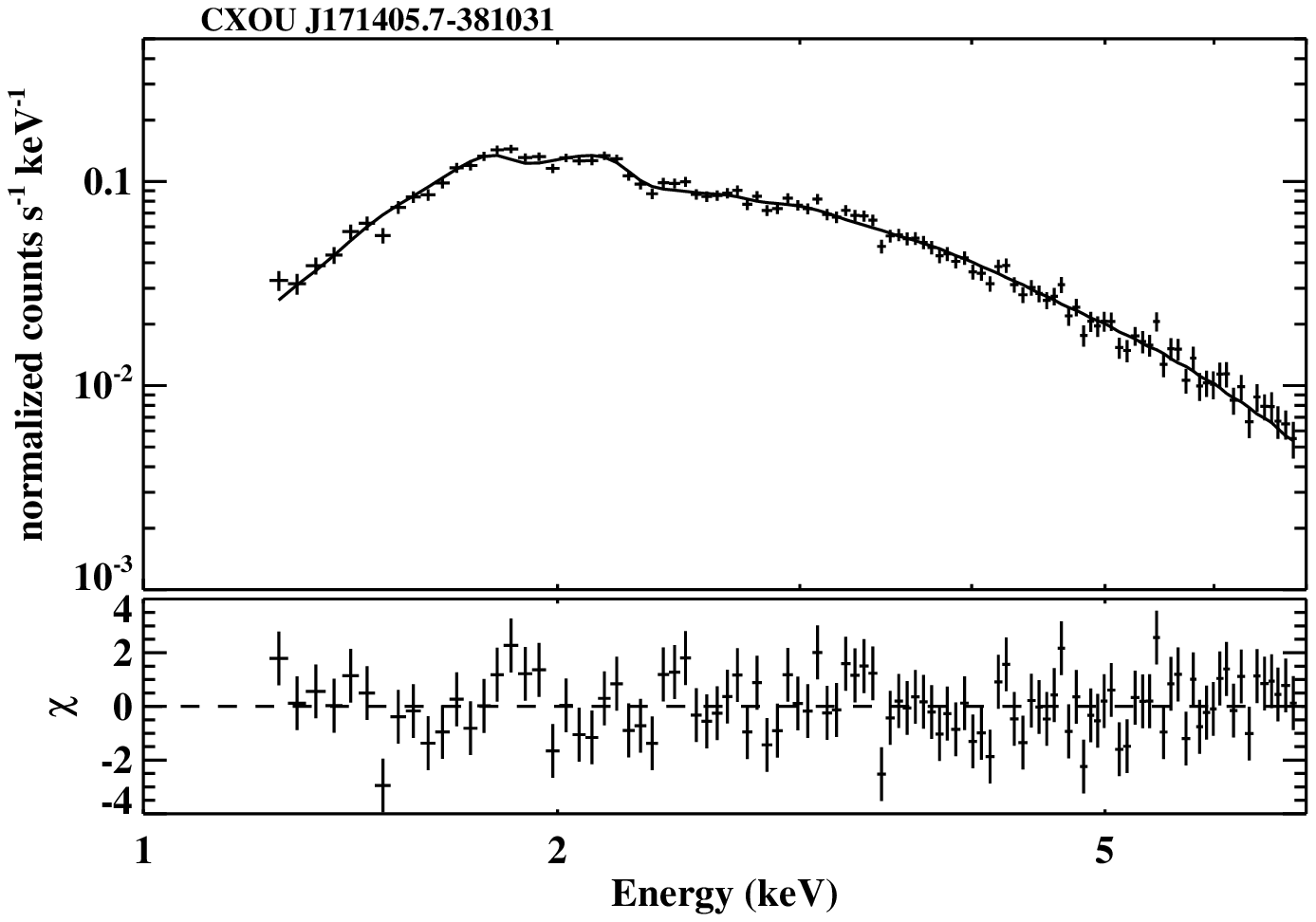}
\includegraphics[scale=0.5]{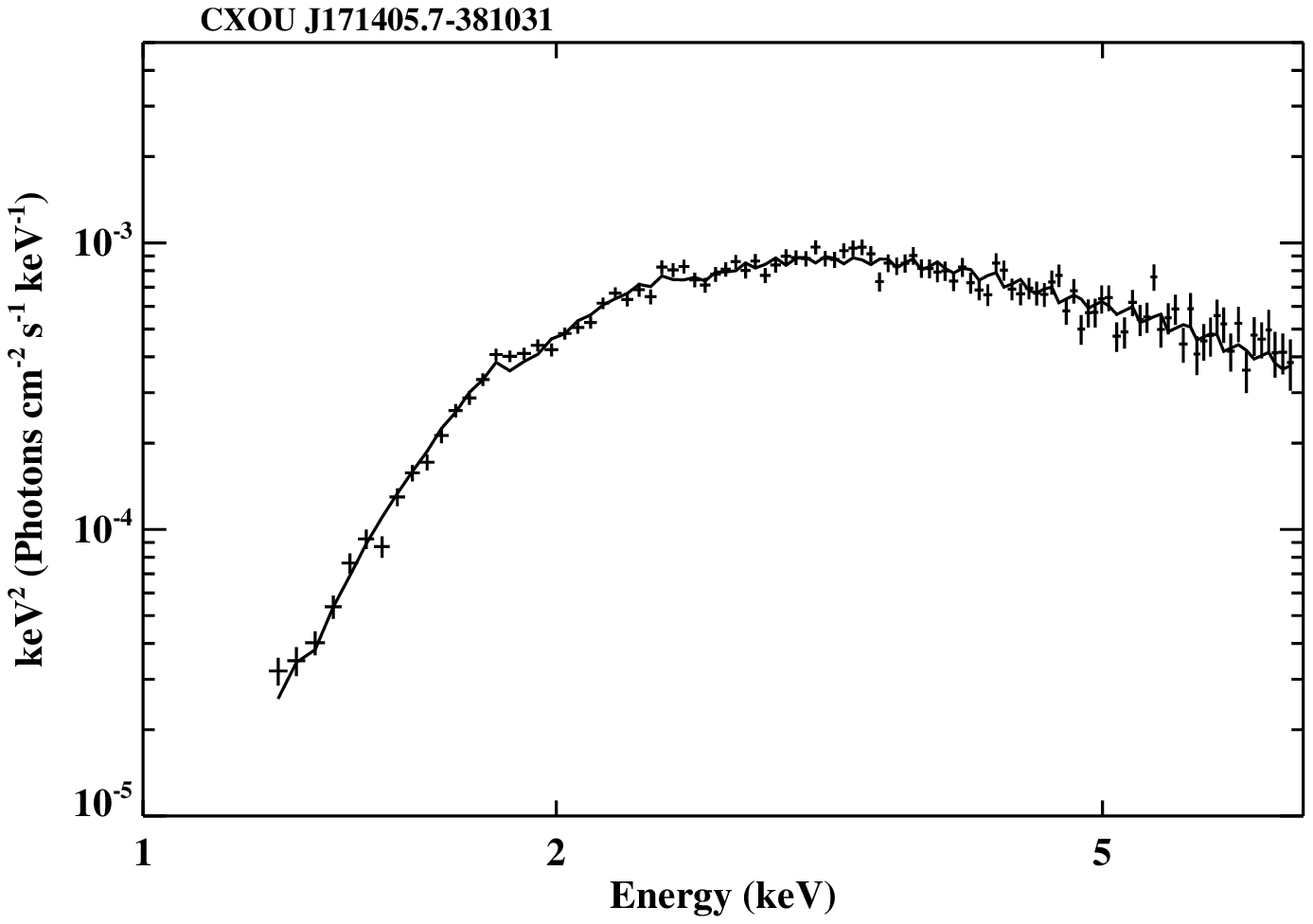}
\caption{Same as Figure \ref{spec_1} but for \bcxouj1714. \label{spec_2}}
\end{center}
\end{figure*}

\begin{figure*}
\begin{center}
\includegraphics[scale=0.5]{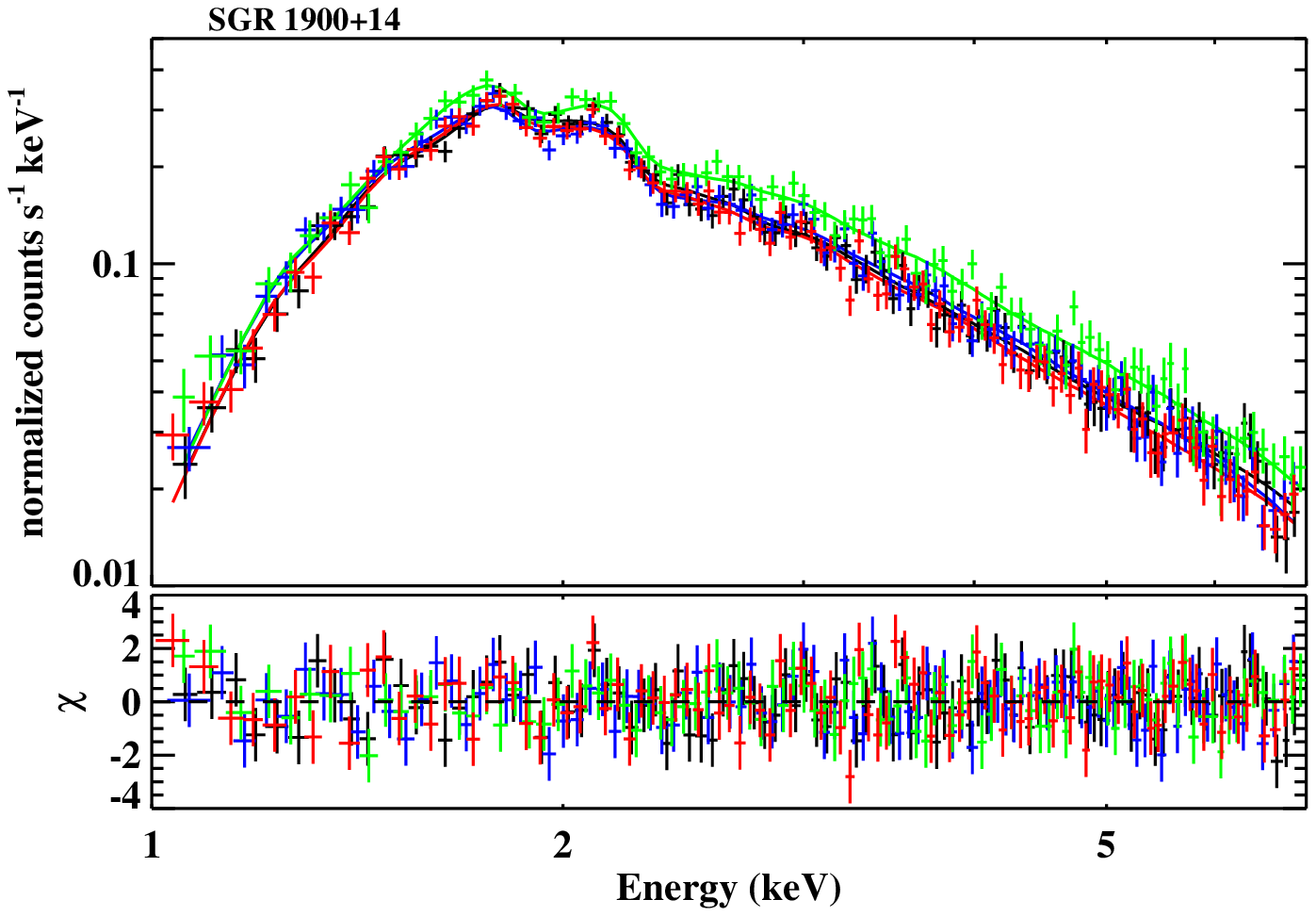}
\includegraphics[scale=0.5]{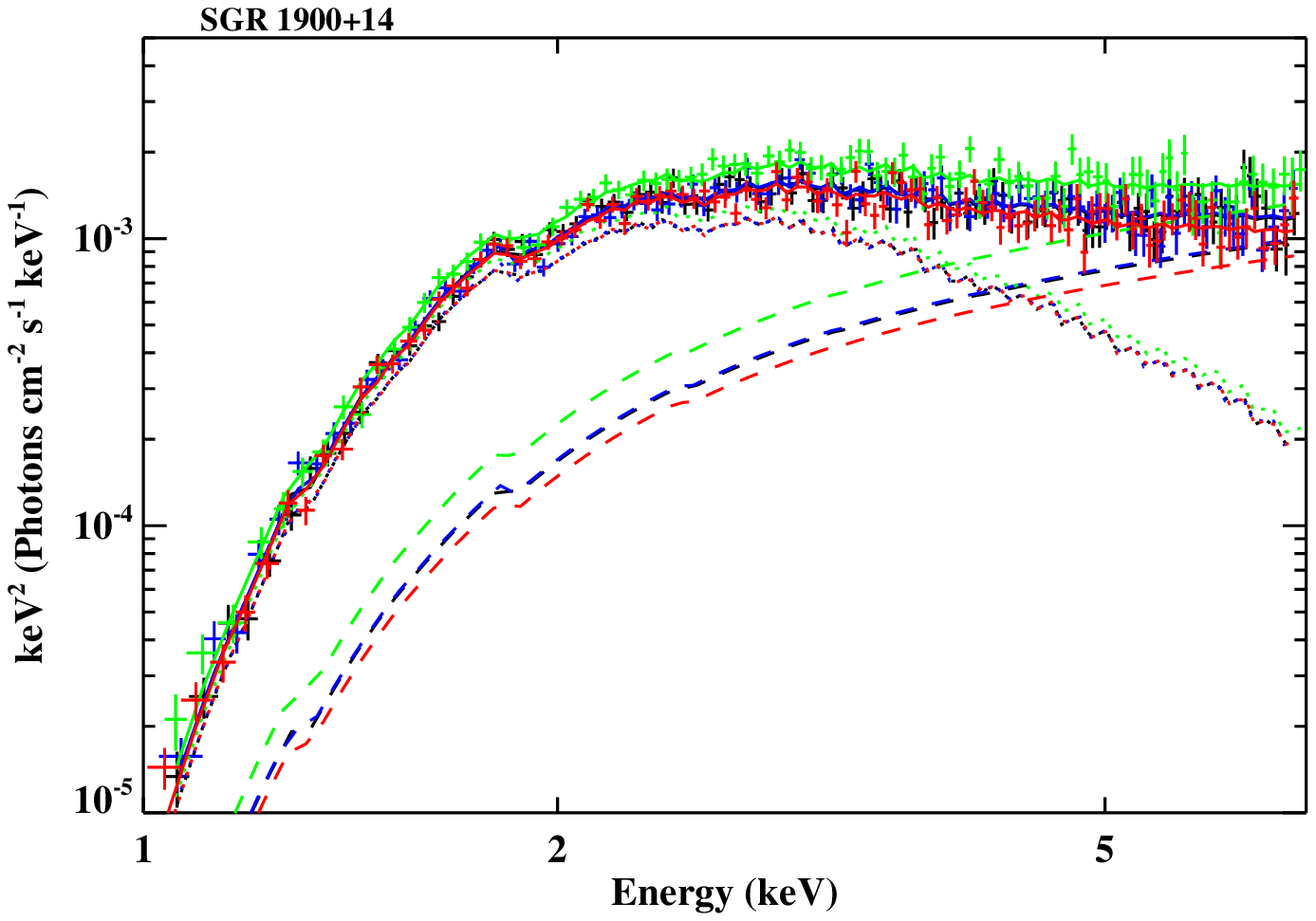}
\caption{Left panel: spectra of \csgr1900 are fitted with the STEMS3D model.
Different colors correspond to the four different epochs of observations and
their fitting residuals are shown in the bottom panel (Table \ref{fits_3}).
Right panel: the corresponding unfolded spectra. \label{spec_3}}
\end{center}
\end{figure*}

\csgr1900 is another source whose broadband spectra (1$-$200 keV) can be
modeled with a BB plus a single PL component. \citet{vrba00} reported that this
source is embedded in a cluster of high-mass stars, at $\sim12-15$ kpc away. We
fit the four {\it XMM-Newton} spectra of \csgr1900 in the 1.0$-$7.0 keV range
in order to eliminate large deviations at the low energy most likely due to the
absorption by the dense gas/dust in the cluster. We find that the surface
temperature remains constant within errors among these observations (see Table
\ref{fits_3}). It is also not likely that the surface magnetic field strength
varies over the course of about 2.5 yr of observational span. We, therefore,
link kT and $B$ in our simultaneous fit so that our modeling would yield a
common value for these two surface parameters. We find that $B = 8.0
\times$10$^{14}$ G, kT $= 0.54$ keV, $\beta \sim 0.11$, and the twist angle is
less constrained. If we further tie $\beta$ and $\Delta\phi$ but allow the PL
and STEMS3D normalizations to vary, we obtain the equivalently good fits
($\chi^{2}/dof = 361.9/392$), with best-fit parameters of $B = 7.81
\times$10$^{14}$ G, $kT = 0.56$ keV, $\Delta\phi = 1.92$, and the $\beta$ hit
the lower limit (0.1) of parameter space (Table \ref{fits_3}). These results
suggest that the physical parameters do not evolve and \csgr1900 is a stable
emitter. The ratio between the PL flux and the STEMS3D flux in 0.5--7.0 keV is
$\sim 30/70$ (Figure \ref{spec_3}).

\begin{deluxetable*}{ccccccccc}
\tabletypesize{\tiny} \tablewidth{0pt} \tablecaption{Spectral parameters of
\csgr1900} \tablehead{\colhead{Group} & \colhead{Obs No.} & \colhead{nH} &
\colhead{kT} & \colhead{$B$} & \colhead{$\Delta\phi$} &
\colhead{$\beta$} & \colhead{Flux} & \colhead{$\chi^2$/dof} \\
\colhead{} & \colhead{} & \colhead{($10^{22}$ cm$^{-2}$)} & \colhead{(keV)} &
\colhead{($10^{14}$ G)} & \colhead{(rad)} & \colhead{} & \colhead{}
&\colhead{}} \startdata

\hline
Group1       & Obs1&  $3.60_{-0.10}^{+0.16}$  & $0.48_{-0.04}^{+0.06}$ & $10.0_{-0.6 }^{     }$ & $0.81_{-0.08}^{+0.12}$ & $0.15_{-0.04}^{+0.03}$ & $3.07$ & 340.7/380\\
     ...     & Obs2&            ...           & $0.43_{-0.18}^{+0.04}$ & $4.79_{-0.64}^{+2.58}$ & $1.90_{-0.89}^{     }$ & $0.20_{-0.03}^{+0.05}$ & $3.08$ &     ...  \\
     ...     & Obs3&            ...           & $0.50_{-0.14}^{+0.08}$ & $8.95_{-0.65}^{+0.53}$ & $1.09_{-0.53}^{+0.87}$ & $0.16_{-0.04}^{+0.06}$ & $3.69$ &     ...  \\
     ...     & Obs4&            ...           & $0.45_{-0.11}^{+0.11}$ & $8.99_{-2.30}^{     }$ & $2.0 _{-1.4 }^{     }$ & $0.14_{     }^{+0.04}$ & $2.93$ &     ...  \\
\hline
Group2       & Obs1&  $3.45_{-0.11}^{+0.12}$  & $0.54_{-0.03}^{+0.02}$ & $8.00_{-0.28}^{+0.28}$ & $1.86_{-0.10}^{+0.11}$ & $0.10_{     }^{+0.02}$ & $3.07$ & 350.3/386\\
     ...     & Obs2&            ...           &          ...           &          ...           & $1.58_{-0.05}^{+0.06}$ & $0.11_{     }^{+0.02}$ & $3.09$ &     ...  \\
     ...     & Obs3&            ...           &          ...           &          ...           & $2.0 _{-0.1 }^{     }$ & $0.12_{-0.02}^{+0.02}$ & $3.68$ &     ...  \\
     ...     & Obs4&            ...           &          ...           &          ...           & $1.87_{-0.06}^{+0.08}$ & $0.10_{     }^{+0.02}$ & $2.93$ &     ...  \\
\hline
Group3       & Obs1&  $3.46_{-0.12}^{+0.13}$  & $0.56_{-0.04}^{+0.01}$ & $7.81_{-0.21}^{+0.61}$ & $1.92_{-0.06}^{+0.07}$ & $0.10_{     }^{+0.04}$ & $3.07$ & 361.9/392\\
     ...     & Obs2&            ...           &          ...           &          ...           &          ...           &          ...           & $3.09$ &     ...  \\
     ...     & Obs3&            ...           &          ...           &          ...           &          ...           &          ...           & $3.68$ &     ...  \\
     ...     & Obs4&            ...           &          ...           &          ...           &          ...           &          ...           & $2.93$ &     ...  \\
\enddata
\tablecomments{Flux: 0.5--7.0 keV absorbed flux in units of $10^{-11}$ erg
cm$^{-2}$ s$^{-1}$. All errors are in the 90\% confidence level. Group1: the
only parameter of nH is linked for different observations in the simultaneous
fit. Group2: parameters of nH, kT, and $B$ are linked. Group3: parameters of
nH, kT, $B$, $\Delta\phi$, and $\beta$ are linked.} \label{fits_3}
\end{deluxetable*}

\subsubsection{Variable Magnetars}

\begin{figure*}
\begin{center}
\includegraphics[scale=0.5]{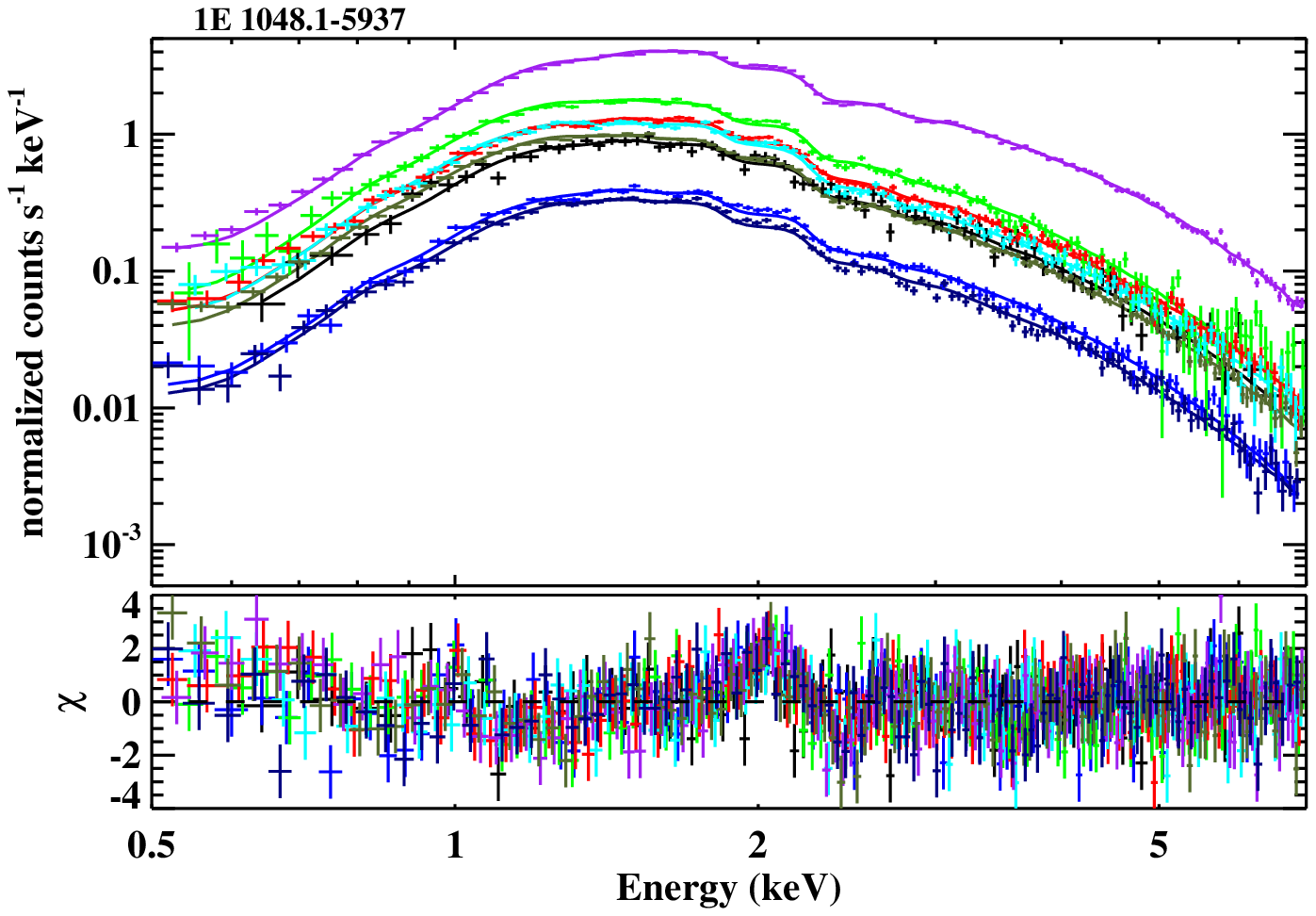}
\includegraphics[scale=0.5]{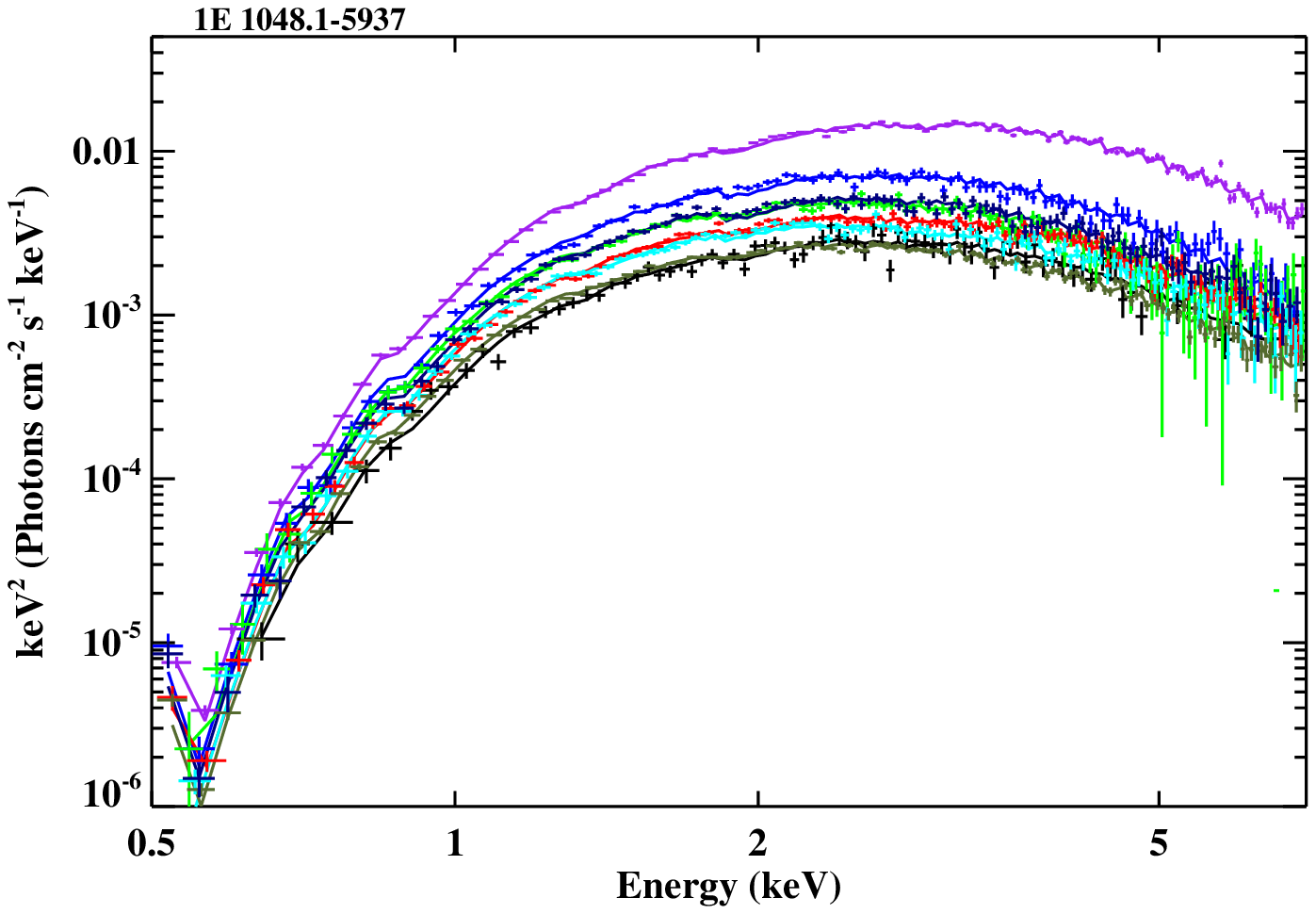}
\includegraphics[scale=0.5]{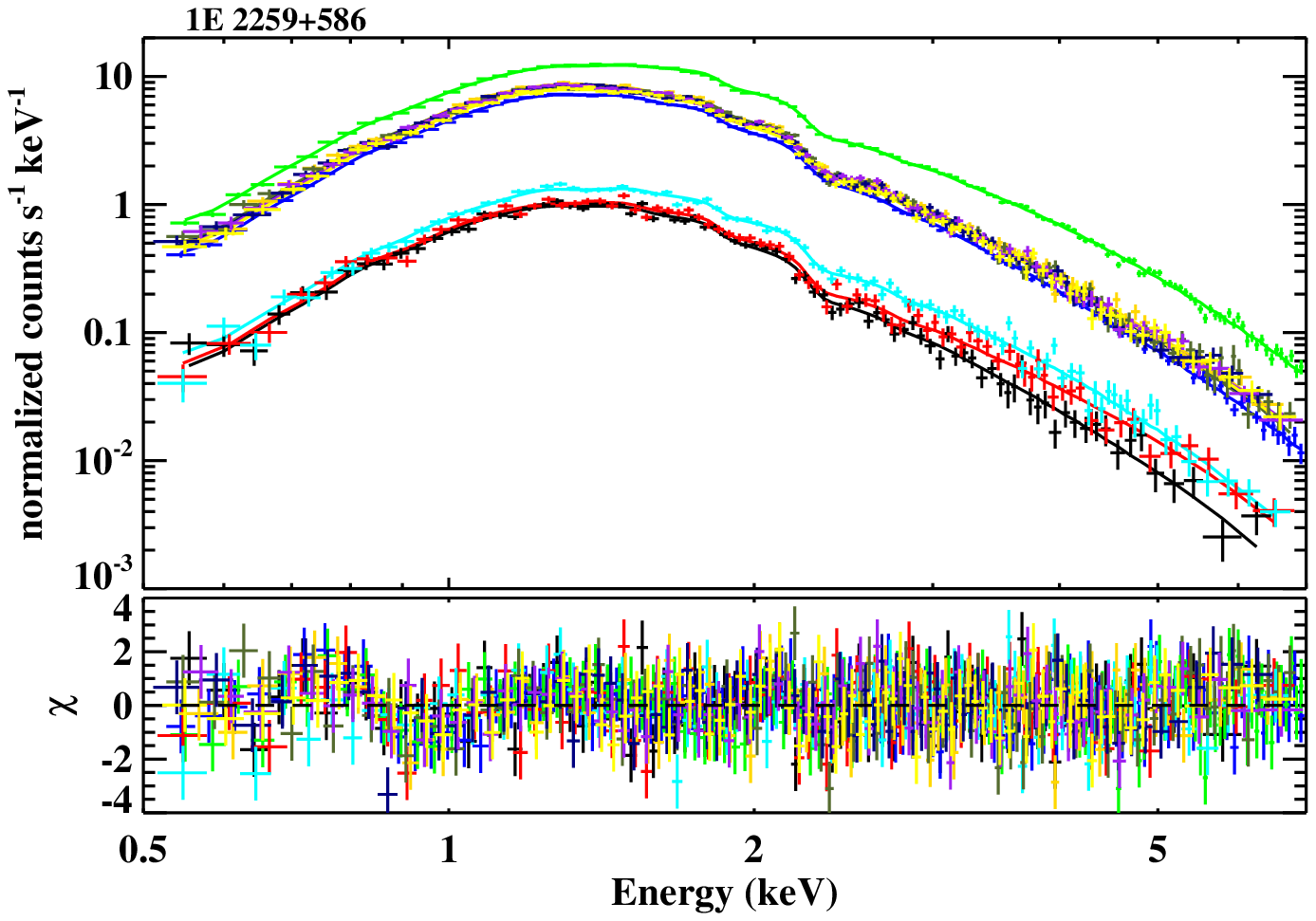}
\includegraphics[scale=0.5]{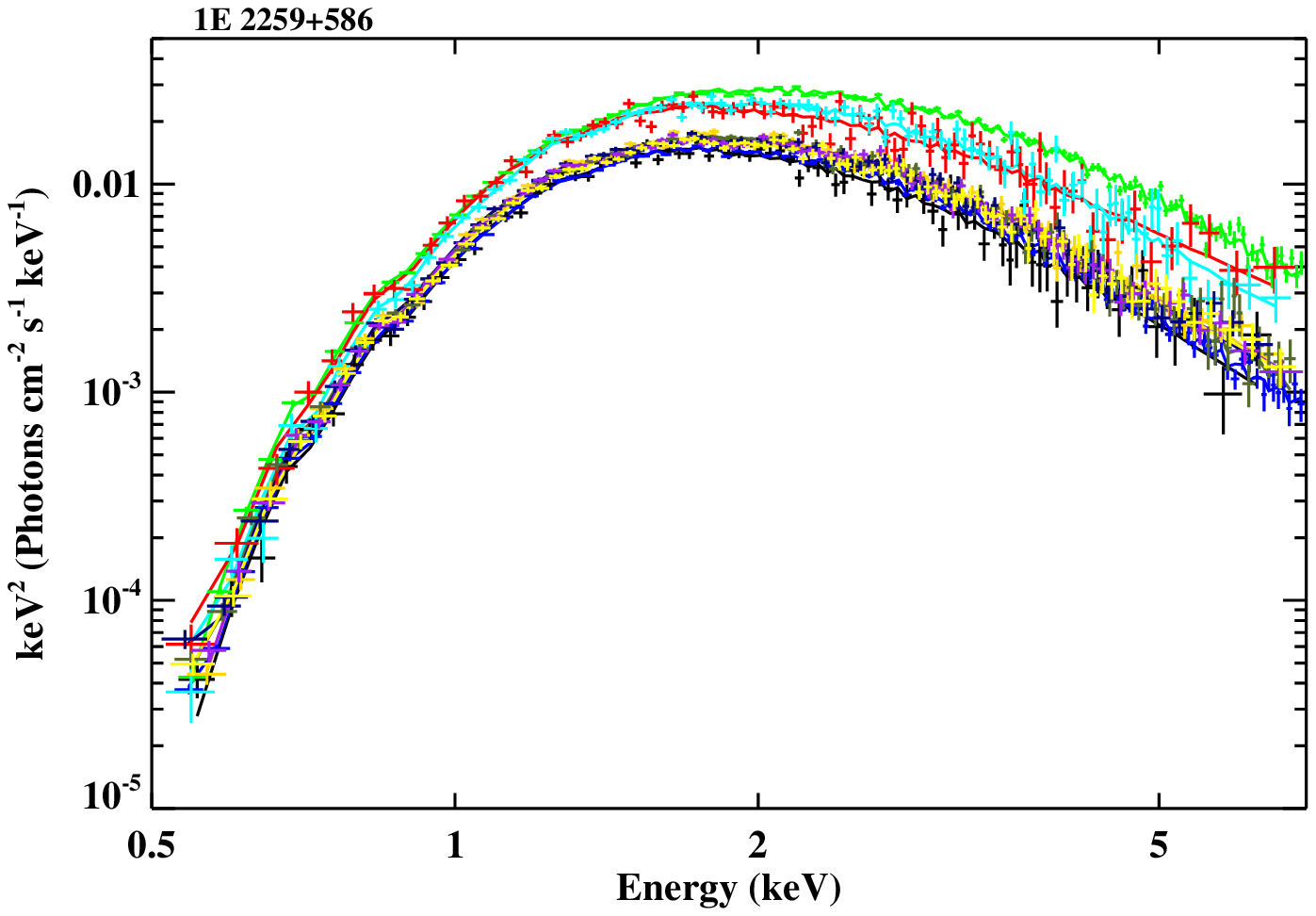}
\caption{Same as Figure \ref{spec_1} but for \axpe1048 and \de2259. \label{var}}
\end{center}
\end{figure*}

\axpe1048 and \de2259 are two bright magnetar sources, whose persistent X-ray
flux varied by a factor of 10 or more over the timescale of months
\citep{dib09, zhu08}. \citet{dib09} performed a comprehensive timing analysis
on $\sim 10$ yr RXTE data of \axpe1048, and showed that all three pulsed flares
were accompanied by temporal events, e.g., significant pulse profile changes
and/or a glitch. The infrared enhancement was also reported at the onset of the
2001 flare \citep{wang02}. We fit all spectra of \axpe1048 with the single
STEMS3D model (Figure \ref{var}) and find that the strength of the surface
magnetic field remains $\sim 2.4\times$10$^{14}$ G (see Group 1 in Table
\ref{fits_4}). By linking the $B$ parameter for all observations, we find that
the kT varies in the range of 0.39$-$0.54 keV, $B = 2.42\times$10$^{14}$ G,
$\Delta\phi > 1.0$, and $\beta$ in the range of 0.12$-$0.16. In order to
provide constraints on the magnetospheric parameters, in addition to $B$, we
link the kT, $\Delta\phi$, and $\beta$ for Obs1 and Obs8, kT and $\beta$ for
Obs5 and Obs7, and kT only for Obs3 and Obs4. This grouping improves the fit
quality slightly ($\chi^{2}/dof = 1113.9/915$, Group 3 in  Table \ref{fits_4}).
Based on these results, we investigated any possible correlations among the
parameters of kT, $\beta$, and 0.5--7.0 keV flux. We find that  kT is
correlated with flux (the Spearman's rank correlation coefficient, $\rho$ =
0.788, the probability of obtaining such a correlation from a random data set,
$P = 0.020$, and $\beta$ is anti-correlated with kT ($\rho/P=-0.672/0.068$), as
well as the flux ($\rho/P=-0.783/0.021$). It is important to note that all of
these correlative behaviors emerge in all three steps of the fitting scheme
presented in Table \ref{fits_4}.

\begin{deluxetable*}{ccccccccc}
\tabletypesize{\tiny} \tablewidth{0pt} \tablecaption{Spectral parameters of
\axpe1048.} \tablehead{\colhead{Group} & \colhead{Obs No.} & \colhead{nH} &
\colhead{kT} & \colhead{$B$} & \colhead{$\Delta\phi$} &
\colhead{$\beta$} & \colhead{Flux} & \colhead{$\chi^2$/dof} \\
\colhead{} & \colhead{} & \colhead{($10^{22}$ cm$^{-2}$)} & \colhead{(keV)} &
\colhead{($10^{14}$ G)} & \colhead{(rad)} & \colhead{} & \colhead{}
&\colhead{}} \startdata

\hline
Group1        & Obs1 &    $1.12_{-0.01}^{+0.02}$ & $0.45_{-0.03}^{+0.03}$ & $2.44_{-0.21}^{+0.16}$ & $1.90_{-1.21}^{     }$ & $0.14_{-0.02}^{+0.04}$ & $0.56$ & 1060.6/902\\
     ...      & Obs2 &             ...           & $0.46_{-0.02}^{+0.02}$ & $2.38_{-0.08}^{+0.08}$ & $1.96_{-0.05}^{     }$ & $0.12_{-0.01}^{+0.01}$ & $1.39$ &     ...  \\
     ...      & Obs3 &             ...           & $0.42_{-0.01}^{+0.01}$ & $2.45_{-0.07}^{+0.07}$ & $1.46_{-0.09}^{+0.31}$ & $0.13_{-0.02}^{+0.02}$ & $0.96$ &     ...  \\
     ...      & Obs4 &             ...           & $0.42_{-0.01}^{+0.01}$ & $2.45_{-0.09}^{+0.08}$ & $1.89_{-0.06}^{+0.08}$ & $0.16_{-0.01}^{+0.01}$ & $0.80$ &     ...  \\
     ...      & Obs5 &             ...           & $0.40_{-0.01}^{+0.01}$ & $2.66_{-0.11}^{+0.11}$ & $2.0 _{-0.1 }^{     }$ & $0.18_{-0.02}^{+0.02}$ & $0.72$ &     ...  \\
     ...      & Obs6 &             ...           & $0.53_{-0.01}^{+0.01}$ & $2.36_{-0.04}^{+0.05}$ & $1.84_{-0.02}^{+0.02}$ & $0.12_{-0.01}^{+0.01}$ & $2.92$ &     ...  \\
     ...      & Obs7 &             ...           & $0.40_{-0.01}^{+0.01}$ & $2.52_{-0.05}^{+0.07}$ & $1.69_{-0.05}^{+0.05}$ & $0.16_{-0.01}^{+0.01}$ & $0.54$ &     ...  \\
     ...      & Obs8 &             ...           & $0.43_{-0.02}^{+0.02}$ & $2.30_{-0.09}^{+0.08}$ & $1.03_{-0.07}^{+0.06}$ & $0.16_{-0.02}^{+0.02}$ & $1.01$ &     ...  \\
\hline
Group2        & Obs1 &    $1.13_{-0.01}^{+0.01}$ & $0.44_{-0.02}^{+0.02}$ & $2.42_{-0.03}^{+0.01}$ & $1.10_{-0.36}^{     }$ & $0.16_{-0.02}^{+0.05}$ & $0.56$ & 1108.3/909\\
     ...      & Obs2 &             ...           & $0.47_{-0.01}^{+0.01}$ &          ...           & $1.96_{-0.05}^{     }$ & $0.12_{-0.01}^{+0.01}$ & $1.39$ &     ...  \\
     ...      & Obs3 &             ...           & $0.42_{-0.01}^{+0.01}$ &          ...           & $1.45_{-0.10}^{+0.09}$ & $0.14_{-0.01}^{+0.01}$ & $0.96$ &     ...  \\
     ...      & Obs4 &             ...           & $0.42_{-0.01}^{+0.01}$ &          ...           & $1.90_{-0.07}^{+0.07}$ & $0.16_{-0.01}^{+0.01}$ & $0.80$ &     ...  \\
     ...      & Obs5 &             ...           & $0.39_{-0.01}^{+0.01}$ &          ...           & $2.0 _{-0.1 }^{     }$ & $0.16_{-0.02}^{+0.02}$ & $0.72$ &     ...  \\
     ...      & Obs6 &             ...           & $0.54_{-0.01}^{+0.01}$ &          ...           & $1.84_{-0.02}^{+0.02}$ & $0.12_{-0.01}^{+0.01}$ & $2.91$ &     ...  \\
     ...      & Obs7 &             ...           & $0.39_{-0.01}^{+0.01}$ &          ...           & $1.54_{-0.08}^{+0.07}$ & $0.16_{-0.01}^{+0.01}$ & $0.54$ &     ...  \\
     ...      & Obs8 &             ...           & $0.44_{-0.01}^{+0.01}$ &          ...           & $1.01_{-0.06}^{+0.07}$ & $0.16_{-0.02}^{+0.02}$ & $1.01$ &     ...  \\
\hline
Group3        & Obs1 &    $1.13_{-0.02}^{+0.02}$ & $0.44_{-0.01}^{+0.01}$ & $2.42_{-0.03}^{+0.03}$ & $1.06_{-0.10}^{+0.10}$ & $0.16_{-0.01}^{+0.02}$ & $0.56$ & 1113.9/915\\
     ...      & Obs2 &             ...           & $0.47_{-0.01}^{+0.01}$ &          ...           & $1.96_{-0.06}^{     }$ & $0.12_{-0.01}^{+0.01}$ & $1.39$ &     ...  \\
     ...      & Obs3 &             ...           & $0.42_{-0.01}^{+0.01}$ &          ...           & $1.45_{-0.04}^{+0.15}$ & $0.13_{-0.01}^{+0.01}$ & $0.96$ &     ...  \\
     ...      & Obs4 &             ...           & $      = obs3        $ &          ...           & $1.90_{-0.64}^{+0.08}$ & $0.16_{-0.01}^{+0.01}$ & $0.80$ &     ...  \\
     ...      & Obs5 &             ...           & $0.39_{-0.01}^{+0.01}$ &          ...           & $2.0 _{-0.1 }^{     }$ & $0.16_{-0.01}^{+0.01}$ & $0.72$ &     ...  \\
     ...      & Obs6 &             ...           & $0.54_{-0.01}^{+0.01}$ &          ...           & $1.84_{-0.02}^{+0.02}$ & $0.12_{-0.01}^{+0.01}$ & $2.91$ &     ...  \\
     ...      & Obs7 &             ...           & $      = obs5        $ &          ...           & $1.45_{-0.12}^{+0.04}$ & $      = obs5        $ & $0.54$ &     ...  \\
     ...      & Obs8 &             ...           & $      = obs1        $ &          ...           & $      = obs1        $ & $      = obs1        $ & $1.01$ &     ...  \\
\enddata
\tablecomments{Flux: 0.5--7.0 keV absorbed flux in units of $10^{-11}$ erg
cm$^{-2}$ s$^{-1}$. All errors are in the 90\% confidence level. Group1: the
only parameter of nH is linked for different observations in the simultaneous
fit. Group2: both parameters of nH, and $B$ are linked. Group3: $\beta$ for
Obs1 and Obs8, kT and $\beta$ for Obs5 and Obs7, and kT only for Obs3 and Obs4
are further linked.} \label{fits_4}
\end{deluxetable*}

The SGR-like bursts and a coincident timing glitch event occurred in the AXP
\de2259 on 2002 June 18 \citep{kaspi03}. The bursts and the glitch (that took
place after Obs2 and before Obs3) were accompanied by the changes in the pulsed
flux/profile and persistent flux/spectrum, providing the strongest evidence of
plastic deformation of the crust \citep{woods04}. Hard-X-ray emission has been
detected in \de2259 by INTEGRAL and Nustar \citep{kuiper06, vogel14}. Since the
hard-X-ray tails contribute negligible flux below 7 keV, in which we are
interested, their soft X-ray spectra are fitted with the single STEMS3D
component (Tables \ref{fits_5} and Figure \ref{var}). It had been suggested by
\citet{zhu08} that the unabsorbed {\it XMM-Newton} flux decay is best described
by the PL and is correlated with the hardness. A simultaneous fit to all 10
observations yields a rather constant surface magnetic field strength. By
linking $B$ and refitting, we obtain $B \sim 6.17 \times$10$^{14}$ G (Group 2
in Table \ref{fits_5}). To better constrain the magnetospheric parameters, we
also merged some successive observations (namely, Obs1/Obs2, Obs4/Obs5,
Obs7/Obs8, and Obs9/Obs10), which have similar surface emission properties (kT)
and are either before or after the bursting episode. We find that the
magnetospheric twist angle rises to its maximum level following the bursting
episode, and declines gradually over a timescale of two to three years. The
particle velocity, on the other hand remains nearly constant around 0.2. Based
on the results of Group 3 in Table \ref{fits_5}, we find a strong correlation
between $\Delta\phi$ and flux with the Spearman's rank correlation coefficient
of $\rho/P=0.942/0.005$. This is an important indication of the
twisted/untwisted magnetosphere scenario, which we will elaborate on in the
Discussion section.

\begin{deluxetable*}{ccccccccc}
\tabletypesize{\tiny} \tablewidth{0pt} \tablecaption{Spectral parameters of
\de2259.}  \tablehead{\colhead{Group} & \colhead{Obs No.} & \colhead{nH} &
\colhead{kT} & \colhead{$B$} & \colhead{$\Delta\phi$} &
\colhead{$\beta$} & \colhead{Flux} & \colhead{$\chi^2$/dof} \\
\colhead{} & \colhead{} & \colhead{($10^{22}$ cm$^{-2}$)} & \colhead{(keV)} &
\colhead{($10^{14}$ G)} & \colhead{(rad)} & \colhead{} & \colhead{}
&\colhead{}} \startdata

\hline
Group1        & Obs1 &  $0.82_{-0.01}^{+0.01}$ & $0.39_{-0.05}^{+0.03}$ & $5.69_{-0.43}^{+0.88}$ & $1.34_{-0.59}^{     }$ & $0.18_{-0.02}^{+0.02}$ & $2.52$ & 871.1/928\\
     ...      & Obs2 &            ...          & $0.35_{-0.01}^{+0.01}$ & $6.01_{-0.21}^{+0.26}$ & $1.45_{-0.05}^{+0.04}$ & $0.20_{-0.01}^{+0.01}$ & $2.60$ &     ...  \\
     ...      & Obs3 &            ...          & $0.33_{-0.01}^{+0.01}$ & $9.61_{-0.28}^{+0.30}$ & $1.97_{-0.22}^{     }$ & $0.22_{-0.01}^{+0.01}$ & $5.62$ &     ...  \\
     ...      & Obs4 &            ...          & $0.36_{-0.05}^{+0.08}$ & $6.14_{-1.50}^{+0.94}$ & $1.10_{-0.26}^{+0.46}$ & $0.23_{-0.01}^{+0.01}$ & $4.46$ &     ...  \\
     ...      & Obs5 &            ...          & $0.42_{-0.03}^{+0.02}$ & $5.79_{-0.41}^{+0.37}$ & $2.0 _{-0.4 }^{     }$ & $0.19_{-0.02}^{+0.02}$ & $4.59$ &     ...  \\
     ...      & Obs6 &            ...          & $0.38_{-0.01}^{+0.02}$ & $5.03_{-0.11}^{+0.25}$ & $1.90_{-0.06}^{+0.07}$ & $0.20_{-0.01}^{+0.01}$ & $3.03$ &     ...  \\
     ...      & Obs7 &            ...          & $0.33_{-0.02}^{+0.03}$ & $7.82_{-0.75}^{+1.01}$ & $1.19_{-0.30}^{+0.12}$ & $0.20_{-0.01}^{+0.01}$ & $2.96$ &     ...  \\
     ...      & Obs8 &            ...          & $0.37_{-0.02}^{+0.02}$ & $5.01_{-0.13}^{+0.81}$ & $1.91_{-0.11}^{     }$ & $0.20_{-0.01}^{+0.01}$ & $2.96$ &     ...  \\
     ...      & Obs9 &            ...          & $0.37_{-0.04}^{+0.02}$ & $5.38_{-0.53}^{+0.21}$ & $1.60_{-0.06}^{     }$ & $0.20_{-0.01}^{+0.01}$ & $2.94$ &     ...  \\
     ...      & Obs10&            ...          & $0.36_{-0.02}^{+0.04}$ & $5.87_{-0.68}^{+0.51}$ & $1.46_{-0.24}^{+0.23}$ & $0.21_{-0.01}^{+0.01}$ & $2.87$ &     ...  \\
\hline
Group2        & Obs1 &  $0.79_{-0.01}^{+0.01}$ & $0.41_{-0.05}^{+0.02}$ & $6.17_{-0.13}^{+0.16}$ & $1.07_{-0.36}^{+0.48}$ & $0.18_{-0.02}^{+0.03}$ & $2.52$ & 930.0/937\\
     ...      & Obs2 &            ...          & $0.37_{-0.01}^{+0.01}$ &          ...           & $1.25_{-0.11}^{+0.04}$ & $0.20_{-0.01}^{+0.01}$ & $2.60$ &     ...  \\
     ...      & Obs3 &            ...          & $0.40_{-0.01}^{+0.01}$ &          ...           & $2.0 _{-0.1 }^{     }$ & $0.23_{-0.01}^{+0.01}$ & $5.63$ &     ...  \\
     ...      & Obs4 &            ...          & $0.38_{-0.05}^{+0.02}$ &          ...           & $1.10_{-0.12}^{+0.30}$ & $0.23_{-0.01}^{+0.01}$ & $4.46$ &     ...  \\
     ...      & Obs5 &            ...          & $0.43_{-0.02}^{+0.02}$ &          ...           & $2.0 _{-0.3 }^{     }$ & $0.18_{-0.02}^{+0.02}$ & $4.59$ &     ...  \\
     ...      & Obs6 &            ...          & $0.37_{-0.02}^{+0.01}$ &          ...           & $1.30_{-0.09}^{+0.09}$ & $0.20_{-0.01}^{+0.01}$ & $3.03$ &     ...  \\
     ...      & Obs7 &            ...          & $0.36_{-0.01}^{+0.02}$ &          ...           & $1.45_{-0.26}^{+0.13}$ & $0.20_{-0.01}^{+0.01}$ & $2.96$ &     ...  \\
     ...      & Obs8 &            ...          & $0.35_{-0.01}^{+0.02}$ &          ...           & $1.50_{-0.26}^{+0.13}$ & $0.20_{-0.01}^{+0.01}$ & $2.96$ &     ...  \\
     ...      & Obs9 &            ...          & $0.37_{-0.02}^{+0.01}$ &          ...           & $1.24_{-0.08}^{+0.32}$ & $0.20_{-0.01}^{+0.01}$ & $2.94$ &     ...  \\
     ...      & Obs10 &            ...         & $0.37_{-0.02}^{+0.01}$ &          ...           & $1.25_{-0.27}^{+0.25}$ & $0.20_{-0.01}^{+0.01}$ & $2.87$ &     ...  \\
\hline
Group3   & Obs1/Obs2 &  $0.77_{-0.01}^{+0.01}$ & $0.40_{-0.01}^{+0.01}$ & $5.89_{-0.08}^{+0.08}$ & $1.46_{-0.03}^{+0.03}$ & $0.19_{-0.01}^{+0.01}$ & $2.56/2.60$ & 976.4/949\\
     ...      & Obs3 &            ...          & $0.43_{-0.01}^{+0.01}$ &          ...           & $2.0 _{-0.1 }^{     }$ & $0.21_{-0.01}^{+0.01}$ & $5.63$ &     ...  \\
     ... & Obs4/Obs5 &            ...          & $0.44_{-0.01}^{+0.02}$ &          ...           & $1.96_{-0.13}^{     }$ & $0.18_{-0.01}^{+0.01}$ & $4.49/4.57$ &     ...  \\
     ...      & Obs6 &            ...          & $0.41_{-0.01}^{+0.01}$ &          ...           & $1.85_{-0.08}^{+0.07}$ & $0.17_{-0.01}^{+0.01}$ & $3.03$ &     ...  \\
     ... & Obs7/Obs8 &            ...          & $0.39_{-0.02}^{+0.02}$ &          ...           & $1.48_{-0.04}^{+0.08}$ & $0.19_{-0.01}^{+0.01}$ & $2.96/2.97$ &     ...  \\
     ... &Obs9/Obs10 &            ...          & $0.41_{-0.02}^{+0.01}$ &          ...           & $1.23_{-0.07}^{+0.04}$ & $0.20_{-0.01}^{+0.01}$ & $2.94/2.87$ &     ...  \\
\enddata
\tablecomments{Flux: 0.5--7.0 keV absorbed flux in units of $10^{-11}$ erg
cm$^{-2}$ s$^{-1}$. All errors are in the 90\% confidence level. Group1: the
only parameter of nH is linked for different observations in the simultaneous
fit. Group2: both parameters of nH, and $B$ are linked. Group3: three pairs of
successive observations (Obs1/Obs2, Obs7/Obs8, and Obs9/Obs10) are merged.}
\label{fits_5}
\end{deluxetable*}

\subsubsection{Transient Magnetars}

\xtej1810 was discovered in 2003 when it suddenly brightened by about two
orders of magnitude above its quiescent flux \citep{ibrahim04, halpern05}. As
it is the first transient magnetar, this source has been visited with {\it
XMM-Newton} more than 24 times and its spectra are well studied, including
investigations with STEMS \citep{guver07}. Due to the fact that the source
shows no clear nonthermal component in quiescence, we only considered the
observations during the outburst decay before 2005 March 18. There are two
observations performed on 2003 September 8, and their spectral fitting results
are consistent with each other. However, the S/N ratio of the second one (ObsID
= 0161360401) is relatively low (due to its short exposure time of $< 1$ ks)
and is excluded here. The surface temperature declines monotonically (from
$0.52\pm0.01$ to $0.34\pm0.01$ keV) following the outburst of \xtej1810, while
the magnetic field remains around $(2.2-2.6)\times$10$^{14}$ G, which is
consistent with those found using STEMS modeling \citep{guver07}. In the
meantime, the $\Delta\phi$ parameter typically remains around $\sim$ 1.7 and
$\beta$ reaches a lower limit of 0.1 (Table \ref{fits_6}). We also fit all four
spectra with common values of $B$ and $\Delta\phi$, which yields $B = 2.52
\times$10$^{14}$ G and $\Delta\phi = 1.66$.

\begin{deluxetable*}{ccccccccc}
\tabletypesize{\tiny} \tablewidth{0pt} \tablecaption{Spectral fit results of
\xtej1810} \tablehead{\colhead{Group} & \colhead{Obs No.} & \colhead{nH} &
\colhead{kT} & \colhead{$B$} & \colhead{$\Delta\phi$} &
\colhead{$\beta$} & \colhead{Flux} & \colhead{$\chi^2$/dof} \\
\colhead{} & \colhead{} & \colhead{($10^{22}$ cm$^{-2}$)} & \colhead{(keV)} &
\colhead{($10^{14}$ G)} & \colhead{(rad)} & \colhead{} & \colhead{}
&\colhead{}} \startdata

\hline
Group1        &Obs1&    $0.88_{-0.02}^{+0.02}$ & $0.52_{-0.01}^{+0.01}$ & $2.22_{-0.04}^{+0.04}$ & $1.66_{-0.02}^{+0.02}$ & $0.10_{     }^{+0.01}$ & $3.76$ & 559.7/442\\
     ...      &Obs2&             ...           & $0.50_{-0.02}^{+0.01}$ & $2.44_{-0.08}^{+0.06}$ & $1.98_{-0.05}^{     }$ & $0.10_{     }^{+0.01}$ & $2.19$ &     ...  \\
     ...      &Obs3&             ...           & $0.44_{-0.01}^{+0.01}$ & $2.50_{-0.06}^{+0.06}$ & $1.66_{-0.04}^{+0.09}$ & $0.11_{-0.01}^{+0.01}$ & $1.32$ &     ...  \\
     ...      &Obs4&             ...           & $0.34_{-0.01}^{+0.01}$ & $2.61_{-0.03}^{+0.04}$ & $1.64_{-0.03}^{+0.03}$ & $0.10_{-0.01}^{+0.01}$ & $0.55$ &     ...  \\
\hline
Group2        &Obs1&    $0.93_{-0.01}^{+0.02}$ & $0.51_{-0.01}^{+0.01}$ & $2.52_{-0.02}^{+0.02}$ & $1.66_{-0.02}^{+0.03}$ & $0.10_{     }^{+0.01}$ & $3.74$ & 750.7/448\\
     ...      &Obs2&             ...           & $0.49_{-0.01}^{+0.01}$ &          ...           &          ...           & $0.10_{     }^{+0.01}$ & $2.19$ &     ...  \\
     ...      &Obs3&             ...           & $0.42_{-0.01}^{+0.01}$ &          ...           &          ...           & $0.12_{-0.01}^{+0.01}$ & $1.32$ &     ...  \\
     ...      &Obs4&             ...           & $0.32_{-0.01}^{+0.01}$ &          ...           &          ...           & $0.10_{     }^{+0.01}$ & $0.55$ &     ...  \\
\enddata

\tablecomments{Flux: 0.5--7.0 keV absorbed flux in units of $10^{-11}$ erg
cm$^{-2}$ s$^{-1}$. All errors are in the 90\% confidence level. Group1: the
only parameter of nH is linked for different observations in the simultaneous
fit. Group2: parameters of nH, $B$, and $\Delta\phi$ are linked.}
\label{fits_6}
\end{deluxetable*}

\begin{figure*}
\begin{center}
\includegraphics[scale=0.5]{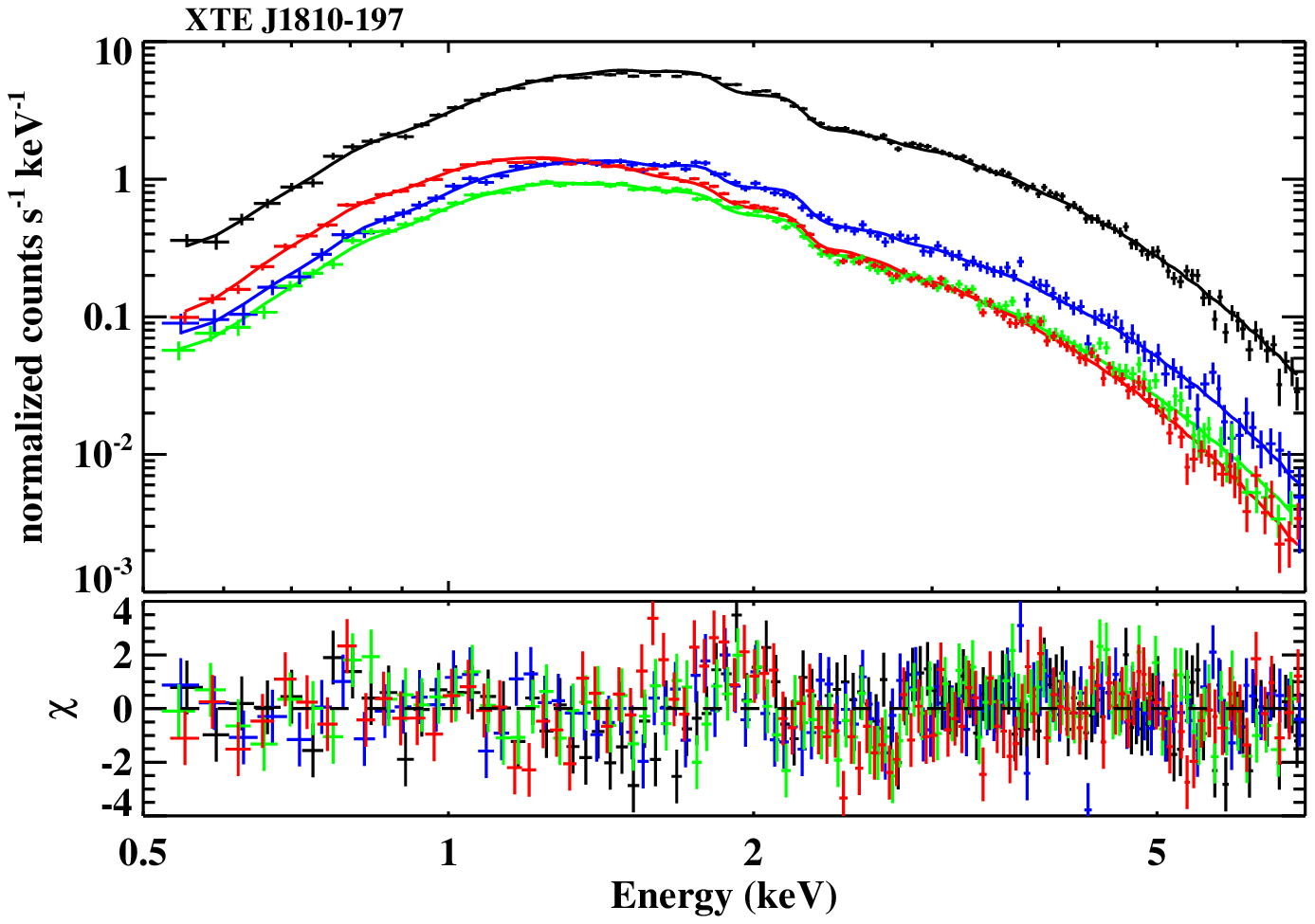}
\includegraphics[scale=0.5]{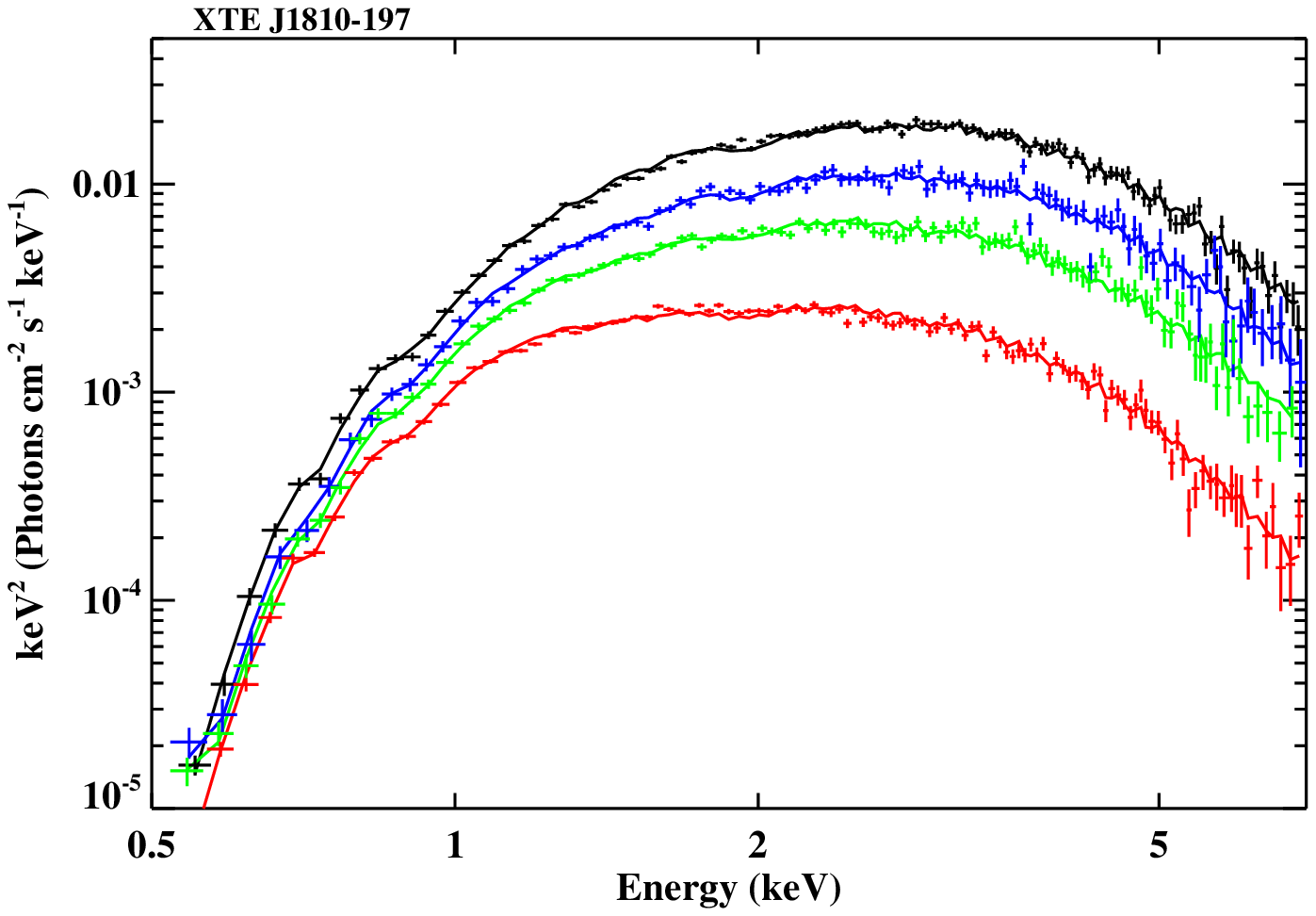}
\includegraphics[scale=0.5]{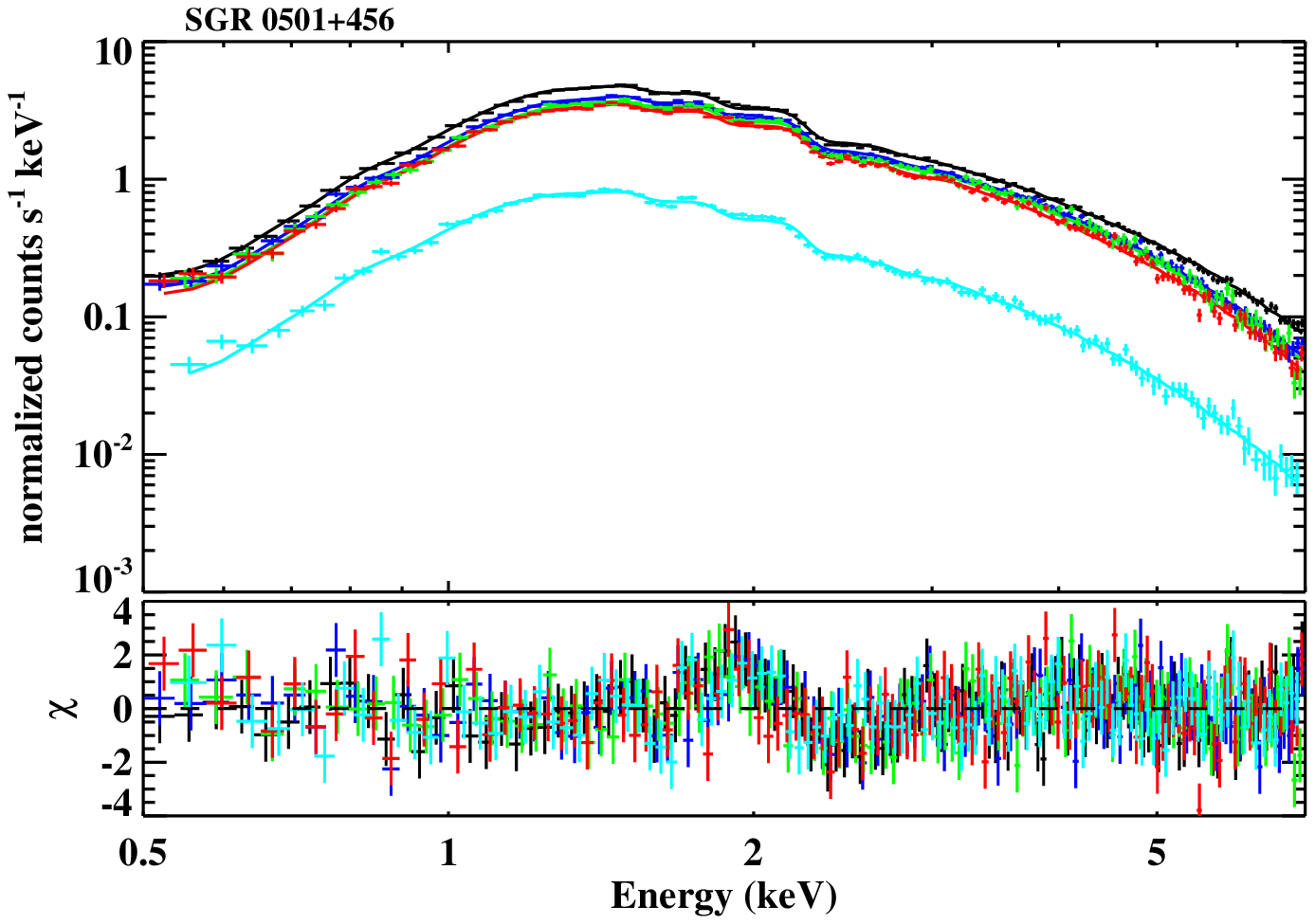}
\includegraphics[scale=0.5]{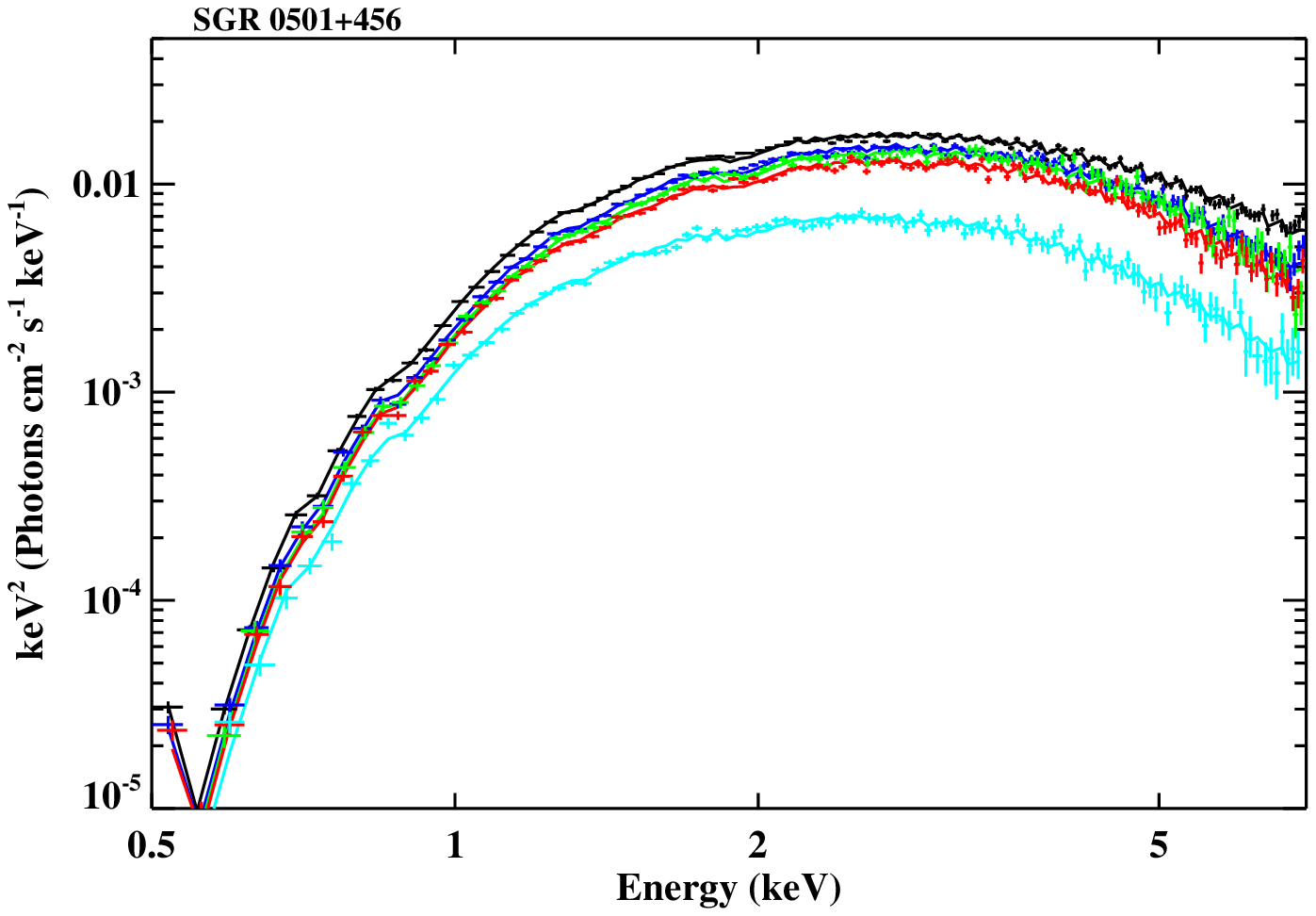}
\includegraphics[scale=0.5]{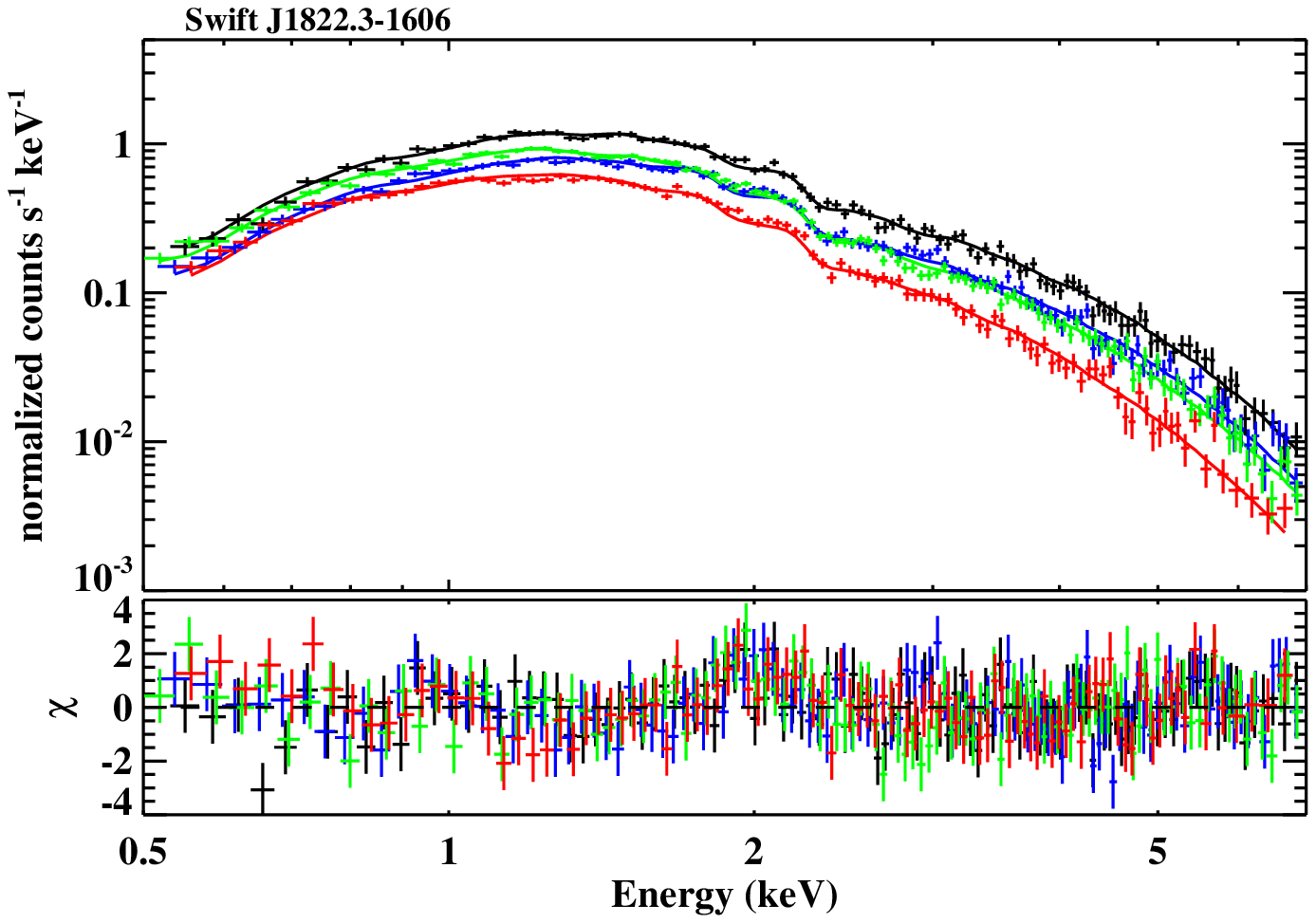}
\includegraphics[scale=0.5]{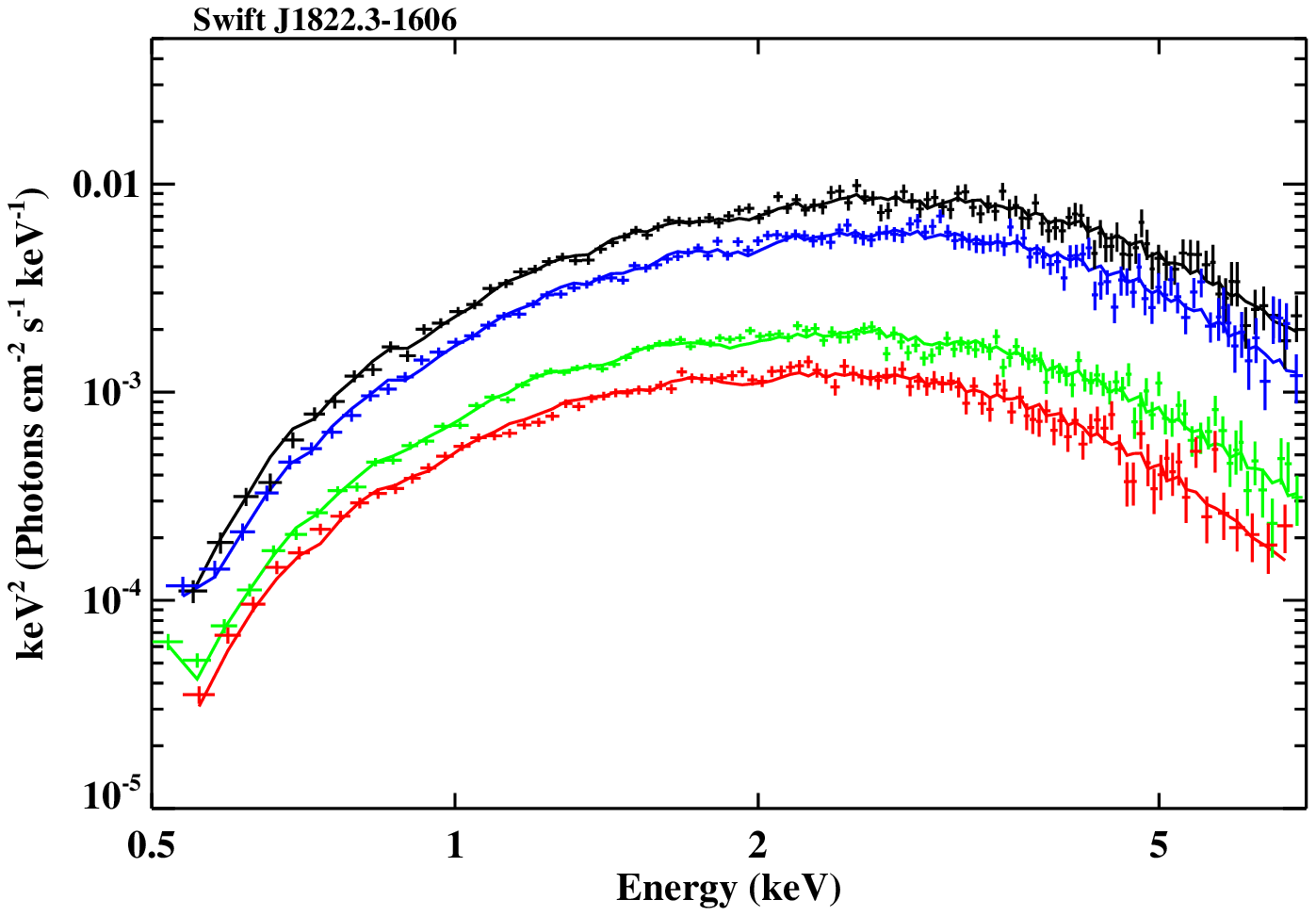}
\caption{Same as Figure \ref{spec_1} but for transients \xtej1810, \bsgr0501,
and \swiftj1822. \label{tran}}
\end{center}
\end{figure*}

\bsgr0501 was discovered because it exhibited a series of short energetic
bursts detected with multiple X-ray missions in 2008 August \citep{gogus10,
kumar10}. The first {\it XMM-Newton} observation of this source took place
while \bsgr0501 was still burst active; therefore, there were a few hundreds of
short bursts in about 45 ks of effective exposure. In order to exclude the
times of these burst events, we constructed the source light curve in a 0.1 s
bin size, and then used a count rate cut-off criterion ($< 55$ cts s$^{-1}$)
\citep{lin12}. \bsgr0501 has also been detected in hard-X-ray band
\citep{enoto10a}, but the hard-X-ray tail does not affect our fitting below 7
keV significantly. The soft-X-ray spectra are, therefore, fit with the STEMS3D
model alone. We find that the surface magnetic field strength remains constant
within errors (see Group 1 in Table \ref{fits_7}). We refit all five spectra
simultaneously to have a common value of $B$, and obtained it as $B = 2.35
\times$10$^{14}$ G. In this case, we obtain that the speed of magnetospheric
electrons declined from 0.2 to 0.15, while the twist angle showed a marginal
evidence of variation between 1.5 and 1.7 (Group 2 in Table \ref{fits_7}).
Among all of the evolving parameters, we find a significant correlation between
only $\beta$ and the flux, with $\rho/P=1/0$.

\begin{deluxetable*}{ccccccccc}
\tabletypesize{\tiny} \tablewidth{0pt} \tablecaption{Spectral fit results of
\bsgr0501}  \tablehead{\colhead{Group} & \colhead{Obs No.} & \colhead{nH} &
\colhead{kT} & \colhead{$B$} & \colhead{$\Delta\phi$} &
\colhead{$\beta$} & \colhead{Flux} & \colhead{$\chi^2$/dof} \\
\colhead{} & \colhead{} & \colhead{($10^{22}$ cm$^{-2}$)} & \colhead{(keV)} &
\colhead{($10^{14}$ G)} & \colhead{(rad)} & \colhead{} & \colhead{}
&\colhead{}} \startdata

\hline
Group1        & Obs1 &    $0.96_{-0.01}^{+0.01}$ & $0.44_{-0.01}^{+0.01}$ & $2.28_{-0.07}^{+0.07}$ & $1.51_{-0.08}^{+0.06}$ & $0.206_{-0.005}^{+0.005}$ & $3.72$ & 600.2/604\\
     ...      & Obs2 &             ...           & $0.49_{-0.01}^{+0.01}$ & $2.33_{-0.07}^{+0.07}$ & $1.53_{-0.08}^{+0.04}$ & $0.159_{-0.008}^{+0.008}$ & $3.13$ &     ...  \\
     ...      & Obs3 &             ...           & $0.50_{-0.02}^{+0.02}$ & $2.34_{-0.10}^{+0.06}$ & $1.71_{-0.06}^{+0.05}$ & $0.150_{-0.021}^{+0.013}$ & $2.94$ &     ...  \\
     ...      & Obs4 &             ...           & $0.50_{-0.02}^{+0.01}$ & $2.41_{-0.08}^{+0.05}$ & $1.66_{-0.14}^{+0.04}$ & $0.143_{-0.016}^{+0.010}$ & $2.65$ &     ...  \\
     ...      & Obs5 &             ...           & $0.44_{-0.02}^{+0.01}$ & $2.41_{-0.09}^{+0.08}$ & $1.65_{-0.30}^{+0.08}$ & $0.144_{-0.011}^{+0.013}$ & $1.44$ &     ...  \\
\hline
Group2        & Obs1 &    $0.96_{-0.01}^{+0.01}$ & $0.45_{-0.01}^{+0.01}$ & $2.35_{-0.04}^{+0.01}$ & $1.50_{-0.06}^{+0.06}$ & $0.203_{-0.010}^{+0.004}$ & $3.72$ & 605.9/608\\
     ...      & Obs2 &             ...           & $0.49_{-0.01}^{+0.01}$ &          ...           & $1.52_{-0.08}^{+0.04}$ & $0.158_{-0.006}^{+0.006}$ & $3.13$ &     ...  \\
     ...      & Obs3 &             ...           & $0.50_{-0.01}^{+0.01}$ &          ...           & $1.71_{-0.06}^{+0.05}$ & $0.150_{-0.019}^{+0.011}$ & $2.94$ &     ...  \\
     ...      & Obs4 &             ...           & $0.49_{-0.01}^{+0.01}$ &          ...           & $1.65_{-0.13}^{+0.04}$ & $0.149_{-0.006}^{+0.006}$ & $2.65$ &     ...  \\
     ...      & Obs5 &             ...           & $0.43_{-0.01}^{+0.01}$ &          ...           & $1.65_{-0.33}^{+0.10}$ & $0.148_{-0.010}^{+0.012}$ & $1.44$ &     ...  \\
\enddata

\tablecomments{Flux: 0.5--7.0 keV absorbed flux in units of $10^{-11}$ erg
cm$^{-2}$ s$^{-1}$. All errors are in the 90\% confidence level. Group1: the
only parameter of nH is linked for different observations in the simultaneous
fit. Group2: both parameters of nH, and $B$ are linked.} \label{fits_7}
\end{deluxetable*}

\swiftj1822 underwent an outburst in 2011 July \citep{livingstone11}. The
timing analysis revealed a spin period of $P = 8.4$ s and a period derivative
of $\dot{P} = 8.3\times10^{-14}$ s s$^{-1}$, which made this source the second
magnetar with a low dipole magnetic field, $B_{\rm timing} \simeq
2.7\times10^{13}$ G \citep{rea12}. Its spectral behavior during the flux decay
resembles those observed from \xtej1810 in the 2003 outburst, whose dipole
magnetic field inferred from spin parameters is $B_{\rm timing} =
2.52\times$10$^{14}$ G \citep{camilo07}. The spectral evolution could be
interpreted as the thermal relaxation in the NS crust \citep{scholz14}. The
long-term monitoring data suggest that the temperature of the thermal component
decreases slowly. Alternatively, the decreasing normalization, which is not
presented in Table \ref{fits_8}, indicates that the emitting region shrinks
during the outburst decay \citep{rea12, scholz14}. In our investigation of this
source with the STEMS3D model, we found that the surface temperature declined
slightly from $0.54\pm0.02$ to $0.44\pm0.02$ keV in the first year of the
outburst, and the surface magnetic field ($B \sim 2.32\times$10$^{14}$ G) and
the electron velocity ($\beta \sim 0.12$) remained constant (Group 1 in Table
\ref{fits_8}). Refitting all four spectra with a linked surface magnetic field
and $\beta$ for all yields a significant fluctuation of $\Delta\phi$ (see Group
2 in Table \ref{fits_8}). We also fit all spectra with the same value of
$\Delta\phi$ and the chi-square increases to $433.5/429$, indicating the a
significant variation of $\Delta\phi$ with an F-test value lower than 0.005.

\begin{deluxetable*}{ccccccccc}
\tabletypesize{\tiny} \tablewidth{0pt} \tablecaption{Spectral fit results of
\swiftj1822}  \tablehead{\colhead{Group} & \colhead{Obs No.} & \colhead{nH} &
\colhead{kT} & \colhead{$B$} & \colhead{$\Delta\phi$} &
\colhead{$\beta$} & \colhead{Flux} & \colhead{$\chi^2$/dof} \\
\colhead{} & \colhead{} & \colhead{($10^{22}$ cm$^{-2}$)} & \colhead{(keV)} &
\colhead{($10^{14}$ G)} & \colhead{(rad)} & \colhead{} & \colhead{}
&\colhead{}} \startdata

\hline
Group1        & Obs1 &    $0.42_{-0.01}^{+0.01}$ & $0.54_{-0.02}^{+0.02}$ & $2.31_{-0.08}^{+0.07}$ & $1.42_{-0.07}^{+0.04}$ & $0.12_{     }^{+0.03}$ & $1.90$ & 415.5/420\\
     ...      & Obs2 &             ...           & $0.54_{-0.02}^{+0.01}$ & $2.33_{-0.04}^{+0.06}$ & $1.52_{-0.06}^{+0.04}$ & $0.11_{     }^{+0.03}$ & $1.31$ &     ...  \\
     ...      & Obs3 &             ...           & $0.45_{-0.02}^{+0.02}$ & $2.34_{-0.08}^{+0.08}$ & $1.80_{-0.03}^{+0.04}$ & $0.13_{-0.01}^{+0.01}$ & $0.43$ &     ...  \\
     ...      & Obs4 &             ...           & $0.44_{-0.02}^{+0.02}$ & $2.35_{-0.09}^{+0.08}$ & $1.69_{-0.04}^{+0.05}$ & $0.12_{-0.01}^{+0.01}$ & $0.28$ &     ...  \\
\hline
Group2        & Obs1 &    $0.42_{-0.01}^{+0.01}$ & $0.54_{-0.01}^{+0.01}$ & $2.32_{-0.04}^{+0.04}$ & $1.42_{-0.07}^{+0.04}$ & $0.12_{-0.01}^{+0.01}$ & $1.91$ & 420.5/426\\
     ...      & Obs2 &             ...           & $0.53_{-0.01}^{+0.01}$ &          ...           & $1.51_{-0.06}^{+0.04}$ &          ...           & $1.31$ &     ...  \\
     ...      & Obs3 &             ...           & $0.46_{-0.01}^{+0.01}$ &          ...           & $1.80_{-0.03}^{+0.04}$ &          ...           & $0.44$ &     ...  \\
     ...      & Obs4 &             ...           & $0.43_{-0.01}^{+0.01}$ &          ...           & $1.69_{-0.04}^{+0.05}$ &          ...           & $0.28$ &     ...  \\
\enddata
\tablecomments{Flux: 0.5--7.0 keV absorbed flux in units of $10^{-11}$ erg
cm$^{-2}$ s$^{-1}$. All errors are in the 90\% confidence level. Group1: the
only parameter of nH is linked for different observations in the simultaneous
fit. Group2: parameters of nH, $B$, and $\beta$ are linked.} \label{fits_8}
\end{deluxetable*}

\begin{figure*}
\begin{center}
\includegraphics[scale=0.5]{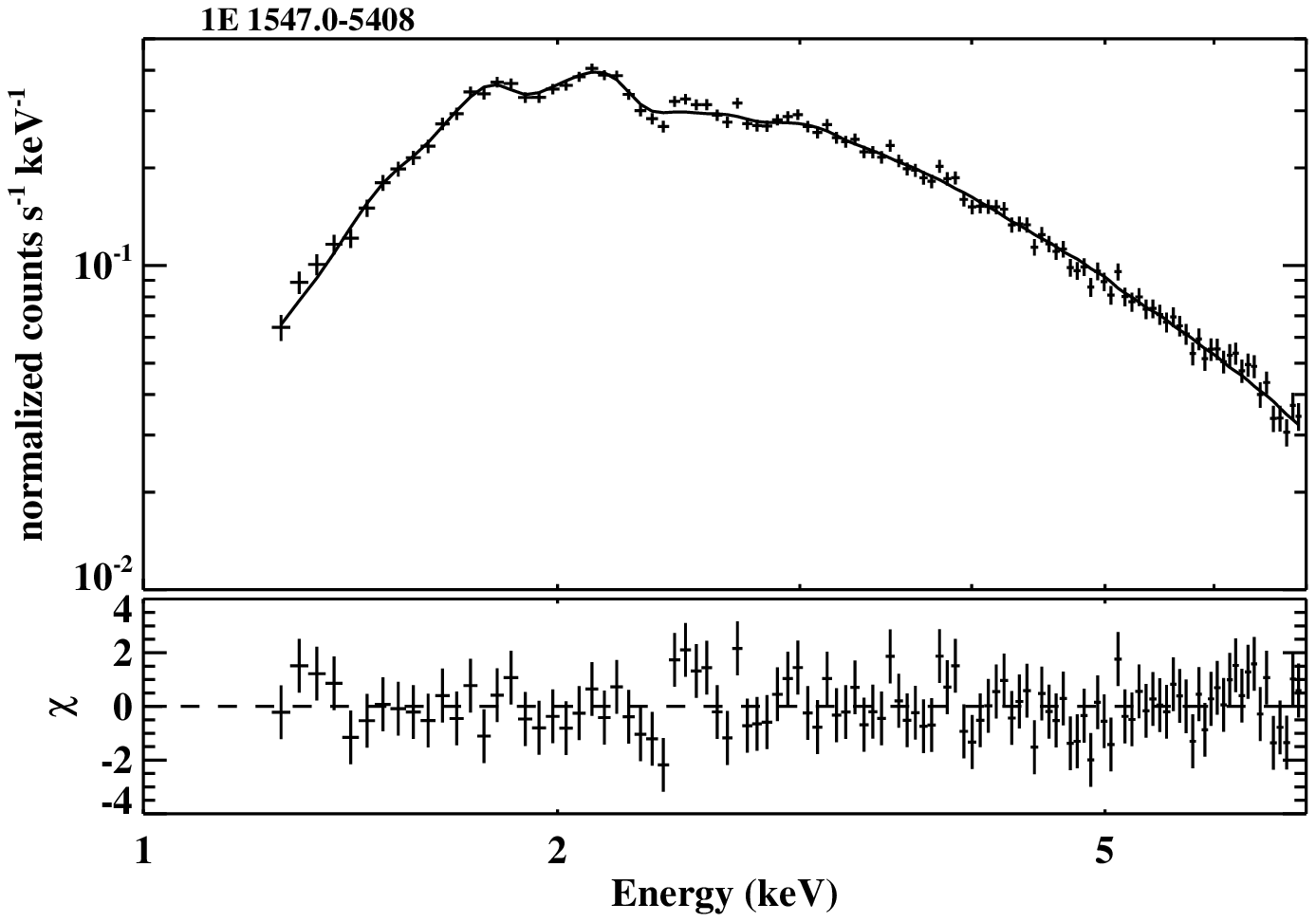}
\includegraphics[scale=0.5]{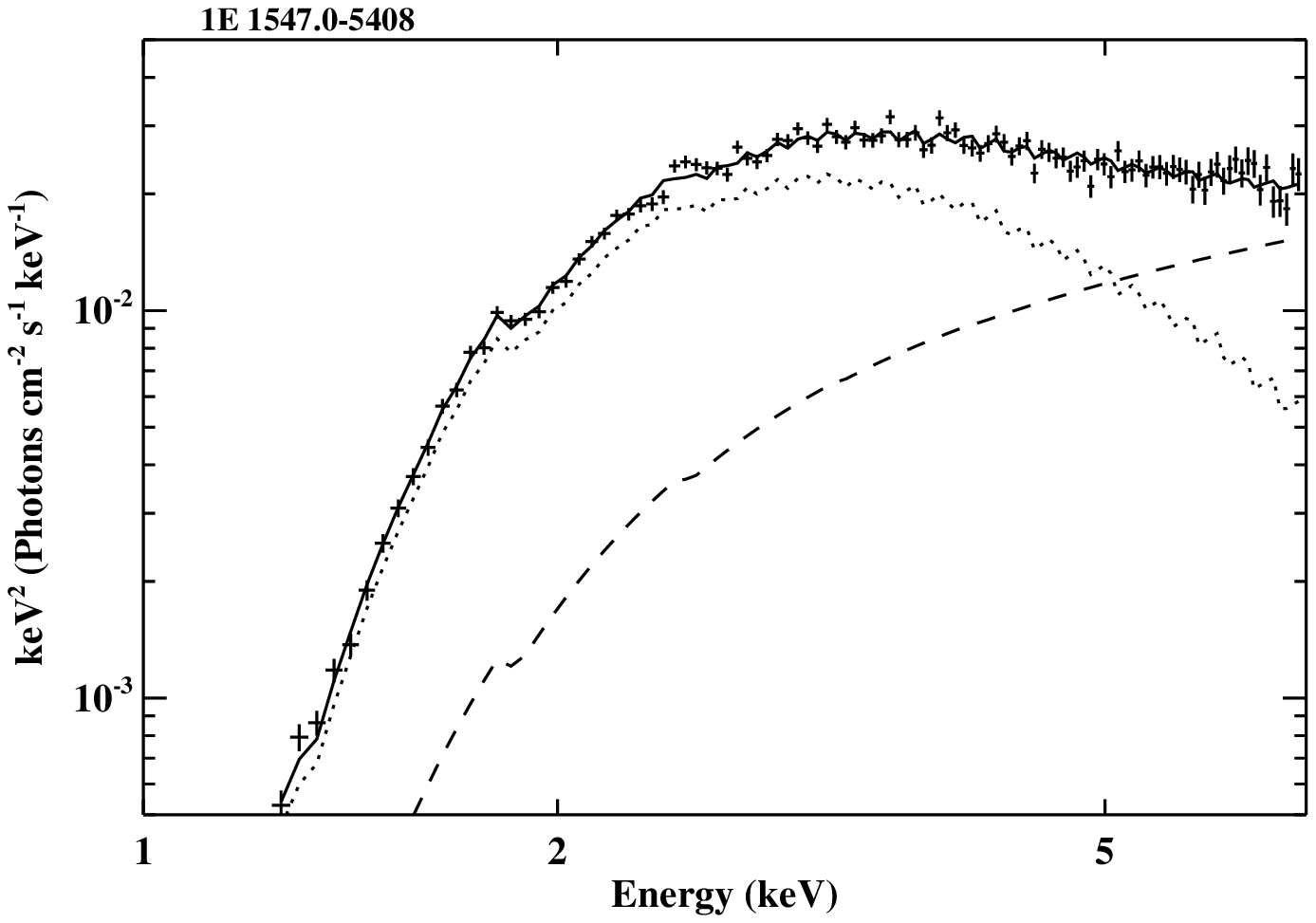}
\caption{Same as Figure \ref{spec_1} but for the observation of \be1547 taken
on 2009 February 3. \label{spec_e1547}}
\end{center}
\end{figure*}

\be1547 was discovered in 1980  \citep{lamb81}, but has shown repeated X-ray
flux enhancements in recent years:  its flux went up by a factor of $> 50$ with
the 2008 October outburst and further increased to levels of $\sim$ 3 orders
brighter than its lowest flux at the onset of its 2009 January outburst
\citep{dib08, kuiper12}. A transient hard-X-ray emission was also recorded with
Suzaku \citep{enoto10b} and INTEGRAL \citep{kuiper12} during the decay of the
2009 outburst. The broadband X-ray spectrum can be described with a BB
component plus a single PL component, and its spectrum evolved significantly in
the soft-X-rays as well as in the hard-X-ray data \citep{kuiper12}. Here, we
only use the {\it XMM-Newton} observation executed on 2009 February 3, which is
one week after the Suzaku observation (on 2009 January 28), and $\Gamma$ of
1.41 is adopted based on the 15--70 keV Suzaku HXD-PIN data \citep{enoto10b}.
The low energy portion of its spectrum is contaminated because the source is
located at the center of SNR G327.24-013 \citep{gelfand07}. Therefore, we again
fit the 1.2--7.0 keV spectrum in order to eliminate large deviations owing to
the contamination of diffuse nonthermal X-rays at the low energy (Figure
\ref{spec_e1547}). We find the best-fit parameters nH $= 4.77_{-0.12}^{+0.13}$
$\times 10^{22}$ cm$^{-2}$, kT $= 0.50_{-0.05}^{+0.03}$ keV, $B =
8.98_{-0.90}^{+0.36} \times$10$^{14}$ G, $\Delta\phi = 1.90_{-0.73}^{+0.09}$,
$\beta = 0.19_{-0.02}^{+0.03}$, absorbed flux in 0.5--7.0 keV $F = 5.14$
$\times 10^{-11}$ erg cm$^{-2}$ s$^{-1}$, and $\chi^{2}/dof = 96.8/102$. Our
modeling reveals that the STEMS3D component dominates the PL component below 7
keV (the flux ratio $\sim 3:1$).

The transient SGR \asgr0418 was discovered on 2009 June 5 when its bursts
triggered the Fermi/GBM, and then was reported as the first magnetar with a low
inferred dipole magnetic field \citet{rea10}. Based on its spectra,
\cite{guver11} argued that the value of the magnetic field at the surface is
$10^{14}$ G based on the STEMS fitting. The strong magnetic field case in this
source was later confirmed by Tiengo et al. (2013). Nevertheless, the STEMS3D
model cannot provide an acceptable fit to the spectrum of \asgr0418 (ObsID =
0610000601). The canonical (BB+PL) model results indicate that the overall flux
is dominated by a very hot thermal component ($kT \sim 0.9$ keV). The STEMS
fitting results show that the particle velocity is $\beta \sim 0.5$ and the
optical depth in the magnetosphere is $\tau \sim 9$. However, as can be seen in
the Figure \ref{optical}, the optical depth in the STEMS3D model is governed by
both $\beta$ and $\Delta\phi$, and cannot be larger than two when the $\beta$
is greater than 0.3.

The AXP \acxouj1647 is located in the young massive stars cluster, Westerlund 1
\citep{muno06}. The source underwent the first intense outburst on 2006
September and the second one on 2011 September \citep{guillermo14}. Measuring
the spin period and period derivative, \citet{an13} suggested that the
spin-down inferred magnetic field strength is less than $7\times10^{13}$ G. We
attempted to fit six bright {\it XMM-Newton} pointing data during 2006 and
2011, and obtained statistically acceptable fits with $\chi^{2}/dof \sim 0.99$.
However, the values of $B$ in both STEMS and STEMS3D modeling peg at the low
limit of $10^{14}$ G; therefore, the obtained parameters are not reliable. We,
therefore, do not include this source for further discussion.

\section{Discussion}

\begin{figure}
\includegraphics[scale=0.4]{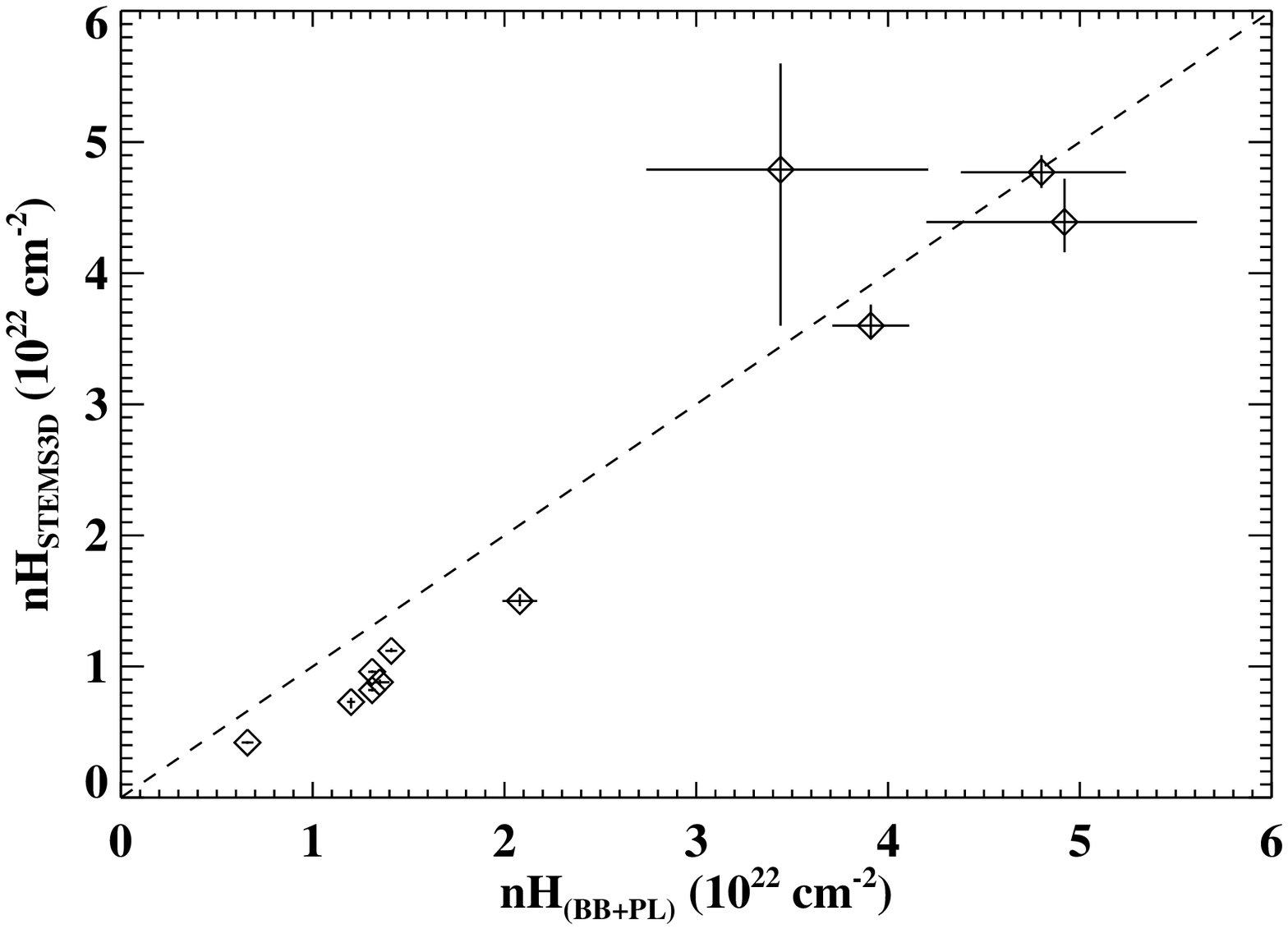}
\includegraphics[scale=0.4]{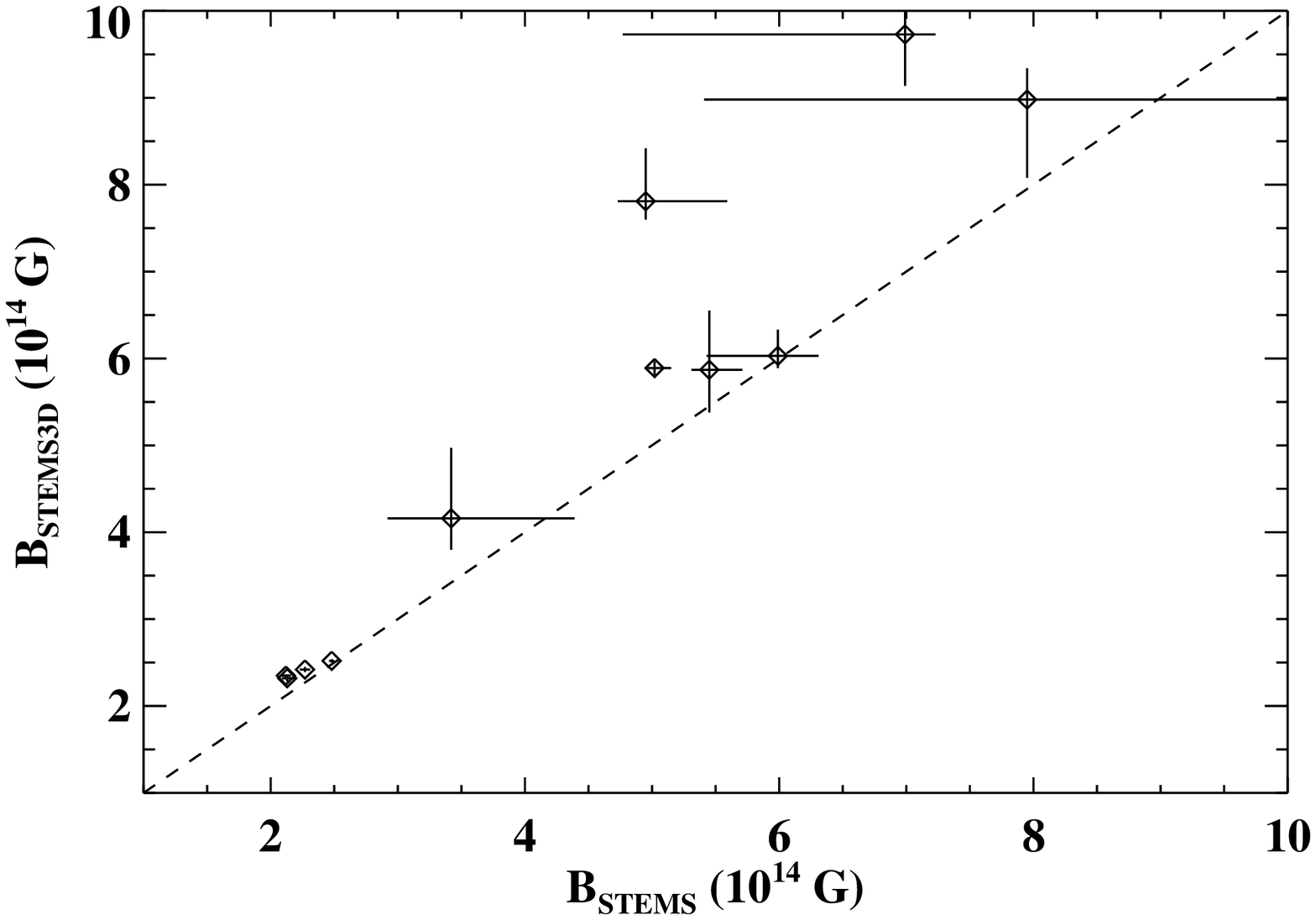} \caption{Upper panel:
nH$_{\rm STEMS3D}$ vs. nH$_{\rm (BB+PL)}$. Bottom panel: $B_{\rm STEMS3D}$ vs.
$B_{\rm STEMS}$. \label{nh_bb}}
\end{figure}

\begin{deluxetable}{cccc}
\tabletypesize{\small} \tablewidth{0pt} \tablecaption{Log of the Magnetic
Field} \tablehead{\colhead{Source} & \colhead{$B_{\rm timing}$} &
\colhead{$B_{\rm STEMS}$} & \colhead{$B_{\rm STEMS3D}$}  \\
\colhead{} & \colhead{($10^{14}$ G)} & \colhead{($10^{14}$ G)} &
\colhead{($10^{14}$ G)}} \startdata

\hline
\axpu0142     & 1.3$^{a}$ & $5.45_{-0.14}^{+0.26}$ & $5.87_{-0.49}^{+0.68}$  \\

\rxsj1708     & 4.6$^{b}$ & $5.99_{-0.56}^{+0.32}$ & $6.03_{-0.14}^{+0.30}$ \\

\ce1841       &  6.9$^{b}$ & $3.42_{-0.50}^{+0.97}$ & $4.16_{-0.36}^{+0.81}$ \\

\bcxouj1714   & 5.0$^{c}$ & $6.99_{-2.22}^{+0.24}$ & $9.73_{-0.59}^{     }$ \\

\csgr1900     & 7.0$^{d}$ & $4.95_{-0.22}^{+0.64}$ & $7.81_{-0.21}^{+0.61}$ \\

\axpe1048       & 3.9$^{e}$ &  $2.27_{-0.04}^{+0.04}$ & $2.42_{-0.03}^{+0.03}$ \\

\de2259       & 0.59$^{f}$ & $5.02_{-0.07}^{+0.13}$ & $5.89_{-0.08}^{+0.08}$ \\

\xtej1810     & 2.1$^{g}$ &  $2.48_{-0.02}^{+0.02}$ & $2.52_{-0.02}^{+0.02}$ \\

\bsgr0501     & 1.9$^{h}$ &  $2.12_{-0.04}^{+0.04}$ & $2.35_{-0.04}^{+0.01}$ \\

\swiftj1822   & 0.135$^{i}$ &  $2.13_{-0.04}^{+0.06}$ & $2.32_{-0.04}^{+0.04}$ \\

\be1547       &  3.2$^{j}$ & $7.95_{-2.54}^{     }$ & $8.98_{-0.90}^{+0.36}$ \\


\enddata

\tablecomments{Period derivatives of magnetars are highly variable, and the
derived $B_{\rm timing}$ could be significantly different from the values
listed here. \\ (a) \citealt{dib07}; (b) \citealt{dib08}; (c) \citealt{sato10};
(d) \citealt{mereghetti06}; (e) \citealt{dib09}; (f) \citealt{gavriil02}; (g)
\citealt{camilo07}; (h) \citealt{gogus10}; (i) \citealt{scholz14};  (j)
\citealt{dib12}.  \label{mag_log}}
\end{deluxetable}

In this paper, we developed a physically motivated numerical model to account
for the magnetar surface emission and magnetospheric scattering in three
dimensions. We applied our model to the X-ray spectra of 13 magnetars,
including persistent X-ray emitters as well as transient sources. We find that
most spectra analyzed here were adequately described with the STEMS3D model,
yielding important physical parameters, such as the surface magnetic field
strengths and degree of twisting in the magnetosphere. To compare with our
results, we also modeled all magnetar spectra with the STEMS model, which
employs the same atmospheric emission properties, but incorporates an only 1D
scattering scheme on the emergent radiation \citep{guver07}. We find that the
surface magnetic field strengths obtained by fitting the STEMS3D model are
generally slightly higher, but are still in agreement within errors with those
obtained with the STEMS model (see the lower panel of Figure \ref{nh_bb}).

Dipole magnetic fields inferred from spin parameters ($B_{\rm timing}$) in 7
out of the 11 sources are consistent with the values obtained from the STEMS3D
model fitting (Table \ref{mag_log}). The STEMS3D model does not provide an
acceptable fit to the spectrum of the first low-$B$ magnetar, \asgr0418, for
reasons as already discussed in \S 4.2.3. While for another two magnetars with
$B_{\rm timing} < 10^{14}$ G, \de2259 and \swiftj1822, the strengths of $B$
given by the STEMS3D model are about one order of magnitude higher than their
dipole magnetic fields. However, taking both twisted magnetospheres
\citep{thompson02, beloborodov09} and particle outflows \citep{tong13b} into
account, the expected spin-down torque ($\dot{P}$) will be larger than a pure
dipole case with the same poloidal magnetic field. The conflict between larger
values of $B_{\rm STEMS3D}$ and timing results could be reconciled by either a
small magnetic inclination angle \citep[e.g.,][]{tong12} and/or a more complex
structure of magnetosphere (i.e., higher order multipoles). In a multipolar
field, the strong surface magnetic field decays more rapidly with radius; thus,
a too fast spin-down rate can be avoided \citep{turolla11}.

On the other hand, an important caveat for the inferred dipole field is that,
magnetars are very noisy and their period derivatives can vary significantly
within a short timescale \citep[e.g.,][and references therein,]{archibald15},
leading to fluctuations in $B_{\rm timing}$, which is unphysical. For example,
\citet{livingstone11} measured a spin-down rate of $\dot{P} =
2.54\times10^{-13}$ s s$^{-1}$ using data collected in the first four months
following the 2011 July outburst of \swiftj1822. \citet{rea13} estimated
$\dot{P} = 8.32\times10^{-14}$ s s$^{-1}$ with the data covering 2011 July --
2012 April, and recently \citet{scholz14} took the new {\it Swift} data into
account and reported $\dot{P} = 2.15\times10^{-14}$ s s$^{-1}$. As a
consequence, the corresponding $B_{\rm timing}$ decreased from the previous
value of $4.7\times$10$^{13}$ G to $1.4\times$10$^{13}$ G listed in Table
\ref{mag_log}. The decreasing spin-down rate can be interpreted as either a
sign of wind braking \citep{tong13} or a glitch recovery \citep{scholz14}. Our
spectral investigations with the STEMS3D, however, yields magnetar-like surface
magnetic field strength ($2.3\times$10$^{14}$ G), which is likely the case
given that, at the very least, the source exhibited energetic bursts.

We find that the temperatures of the magnetar surfaces vary between 0.24 and
0.56 keV, and the magnetospheric electrons are non-relativistic ($\beta \leq$
0.2). We investigated multiple {\it XMM-Newton} observations typically spanning
years of five magnetars and find that the strength of the surface magnetic
fields remains nearly constant. We also modeled all magnetar spectra with the
phenomenological (BB+PL) model to better establish the interstellar absorbtion
aspects toward magnetars studied here. It is well known that the simple PL
model increases very rapidly at low energies, and as a consequence, a larger
absorption column density is required to account for the overestimated PL flux,
especially for the steep PL indices \citep{durant06, guver08}. In all sources
except \ce1841, the hydrogen column density obtained from the STEMS3D model
fits (nH$_{\rm STEMS3D}$) is lower than the value derived from the (BB+PL)
model fits (nH$_{\rm (BB+PL)}$; Figure \ref{nh_bb}). We argue that, given the
empirical nature of the (BB+PL) model, the density of interstellar hydrogen
obtained with the STEMS3D is a more reliable indicator.

As already stated, the surface magnetic field values from the STEMS3D model
mostly agree with those obtained with the STEMS model results (Table
\ref{mag_log}). However, the STEMS model makes simplifying assumptions about
the magnetic field geometry, i.e., treating the scattering region as a
plane-parallel slab; therefore, it cannot offer information about the geometry
of the magnetosphere. Alternatively, the 3D self-similar magnetic configuration
is characterized by a free parameter, $\Delta\phi$, in the STEMS3D model. Our
fitting results indicate that the magnetospheres in most magnetars are highly
twisted ($\Delta\phi \ge 1$). It is worth noting that we also included a flat
PL ($\Gamma = 1.32$) component in the fitting for \ce1841, which could be the
reason why it has a large value of $nH_{\rm STEMS3D}$. The derived small
magnetospheric twist angle $\Delta\phi = 0.31_{-0.11}^{+0.80}$ and large $\beta
= 0.28_{-0.06}^{+0.04}$ should be treated with caution, and longer exposure
observations are required to check the magnetospheric parameters of \ce1841 in
the future.

Compared with the STEMS model, the STEMS3D model embodies some other
significant improvements: (1) in the 1D geometry, the magnetic field is fixed
to follow a $\mathbf{B} \propto r^{-3}$ dependence, and the optical depth
$\tau$ is independent of the electron velocity $\beta$. In contrast, the radial
dependence of the magnetic field is consistently obtained from $\Delta\phi$,
while the angle-averaged optical depth $\tau$ is determined by $\beta$ and
$\Delta\phi$. (2) The charged particles are assumed to have a top-hat velocity
distribution in the 1D model, while a more realistic distribution, involving
the effects of bulk motion and also thermal velocity distribution of electrons,
is used in the STEMS3D model. Following the work in \citet{nobili08}, we adopt
the 1D Maxwellian distribution at the electron temperature $T_{\rm e}$, which
is further linked with $\beta$ by assuming equipartition between thermal and
bulk kinetic energy in order to reduce the number of parameters in the model.
(3) In the 3D configuration, the optical depth varies significantly with angles
and vanishes at magnetic poles in the twisted magnetosphere. Thus, it is
possible to detect the surface radiation which is mostly carried by
extraordinary mode photons, if one of the magnetic poles is in our line of
sight. The polarization and the proton cyclotron absorption lines are expected
to vary with the rotational phase. Note that \cite{tiengo13} recently reported
a phase-variable absorption feature observed during the outburst of \asgr0418.

We also aimed to understand the nature of variable and transient magnetars by
investigating their spectral evolution. For this purpose, we analyzed multiple
{\it XMM-Newton} observations of six magnetars. \axpe1048 has been quite active
in the past 10 yr: three X-ray flares were recorded with RXTE in 2001, 2002,
and 2007 March \citep{dib09}. The eight {\it XMM-Newton} observations we used
here are sampled before, after and in between these flares: Obs1 was performed
before the first flare; Obs2-5 were executed between the second and third
flares; Obs6-7 were performed after the third flare; comparing with Obs7, the
source brightened by a factor of two in Obs8 (on 2013 July 22, after the end of
RXTE mission). We find that the surface temperature monotonically decreases
with declining flux from Obs2 to Obs5 and from Obs6 to Obs7 (see Group 1--3 in
Table \ref{fits_4}). This is likely due to cooling of the NS crust that is
heated by the mechanism leading to these flares. However, the magnetospheric
twist angle cannot be constrained in this source.

In the magnetar framework, both the crustal cooling \citep[e.g.,][]{pons12} and
the magnetospheric relaxation \citep[e.g.,][]{bernardini09} models are proposed
for the flux relaxation in different sources. The dense {\it XMM-Newton}
observations on \de2259 provide a good opportunity to study this topic. The
first two data sets are observed before the burst event, and the other eight
observations monitored the flux relaxation of the outburst in the following
three years. The STEMS3D fitting results suggest a large magnetic field $\sim
5.9\times$10$^{14}$ G for the source, and the $\Delta\phi$ suddenly increased
to a maximum from Obs2 to Obs3 then decreased following the flux decays. The
only exception, Obs4, which was executed $\sim 20$ days after the burst, has a
small $\Delta\phi = 1.1$. Note that, Obs4 and Obs5 were performed on the same
day; however, the latter yields a large $\Delta\phi \sim 2$. Such a small
discrepancy between Obs4 and Obs5 was also seen in the (BB+PL) fitting results
\citep{zhu08}, i.e., the flux ratio between PL and BB components varied from
2.6 (Obs4) to 2.1 (Obs5). It is not likely that the magnetosphere could untwist
and retwist in a few days, and the small $\Delta\phi$ derived from Obs4 is
dubious. Group 2 in Table \ref{fits_5} shows a strong correlation between
$\Delta\phi$ and flux with the Spearman's rank correlation coefficient of
$\rho/P=0.661/0.038$. Since the Spearman's rank correlation coefficient does
not consider the parameter uncertainties, it would underestimate the
significance level due to the spectral fluctuations. Therefore, we merge three
pairs of observations (Obs1/Obs2, Obs7/Obs8, and Obs9/Obs10) that have similar
spectral parameters and flux levels, as well as Obs4/Obs5, and the correlation
between $\Delta\phi$ and flux becomes more significant ($\rho/P=0.942/0.005$,
Group 3 in Table \ref{fits_5}). When $\Delta\phi$ is linked, the $\chi^{2}$
increases by more than 51.9, which suggests the significant variation of
$\Delta\phi$ with an F-test probability lower than $2\times10^{-9}$. Therefore,
our finding of an increased twist angle accompanied with the outburst (occurred
between Obs2 and Obs3) and then a decrease following the source decays, agrees
perfectly with the twisted/untwisted magnetosphere scenario \citep{thompson02}.

The STEMS3D fitting results of both \xtej1810 and \swiftj1822 show that the
surface temperature monotonically declines with decreasing luminosity. The
former source had been studied in detail with the STEMS model \citep{guver07}.
On the other hand, it is the first time that we evaluated the surface magnetic
field of \swiftj1822 via continuum-fitting, and obtained a significantly
stronger magnetic field than the dipole magnetic field inferred from the
spin-down rate \citep{scholz14}. It is interesting to note that \xtej1810 and
\swiftj1822 share the similar spectral sequences following their outbursts
\citep{rea12, scholz14}. Nevertheless, the timing analyses infer a typical
magnetar magnetic field for \xtej1810 \citep[$B_{\rm timing} \sim
2.1\times$10$^{14}$ G, ][]{camilo07} while a low value for the latter source
\citep[$B_{\rm timing} \sim 1.4\times$10$^{13}$ G, ][]{scholz14}. Our
spectroscopic measurements suggest a similar magnetic field strength of $B_{\rm
STEMS3D} \sim 2.3\times$10$^{14}$ G for these two transients. The monotonic
decline of surface temperatures during the outburst decays in both \xtej1810
and \swiftj1822 are remarkable evidence of crustal cooling
\citep{guver07,scholz14}. Additionally, we also find a significant variation of
$\Delta\phi$ in \swiftj1822 (with a confidence of 99.5\%). We also find a
decline of the magnetospheric electron velocity $\beta$ in the first $\sim 40$
days of the 2008 outburst of \bsgr0501 (Table \ref{fits_7}), implying the
energy dissipation of charged particles, probably by the radiative drag.

Different scenarios have been proposed to account for the hard-X-ray emissions
from magnetars. \cite{thompson05} put forward two mechanisms that
soft-$\gamma$-rays either arise from the thermal bremsstrahlung emissions in a
thin surface layer heated by a returning current or the synchrotron emissions
from $e^{\pm}$ pairs \citep[see also][]{beloborodov07}. Alternatively,
\cite{baring07} suggested that the high energy tail was produced by the
resonant up-scattering in the magnetosphere. The 3D Monte Carlo simulations
show that the soft-X-ray photons can be up-scattered to $\sim 100$ keV if the
scattering particles are energetic, i.e., $\beta > 0.6$ and the corresponding
electron temperature $kT_{\rm e} > 100$ keV \citep{nobili08, zane09}. However,
our spectral analysis shows that the $\Delta\phi > 1 $ and the $\beta \sim 0.2$
in most cases, while flatter spectra and stronger pulsations detected at the
hard-X-ray clearly point to a different population of electrons.
\cite{beloborodov13} proposed that e$^{\pm}$ pairs discharge near the neutron
star and create the relativistic outflow that further scatters thermal photons
to high energies. This model provided a satisfactory explanation for the
hard-X-ray luminosity, spectral slopes, and pulsed profiles
\citep[e.g.,][]{hascoet14, vogel14}, and it may also contribute an
non-negligible flux in the soft-X-ray band.

It should be noted that the STEMS3D model still has some limitations, as other
magnetospheric scattering models do (see \citealt{zane09} and
\citealt{beloborodov13}). Both spectral and temporal analyses have demonstrated
that the thermal emission is not uniformly distributed on the surface of
magnetars but are confined in a small region \citep[e.g.,][]{ozel02, albano10}.
However, because the uniform distribution is adopted for the seed photons, the
STEMS3D model cannot provide an acceptable fit to the spectra of magnetars in
quiescence, which lacks nonthermal emission. As the transients return to the
quiescent state, the spectra of the three transients studied in our paper
(i.e., \xtej1810, \bsgr0501, and \swiftj1822) became significantly softer than
those at outburst levels, and the second or even third thermally emitting
surface regions were reported \citep{bernardini09, camero14,scholz14}.
According to the (BB+BB) or (BB+PL) fits, it is found that the thermal emitting
area decreases following the outburst decay. In these cases, the distinctive
spectra would be formed for the following reasons: (i) In the 3D twisted
magnetosphere, the optical depth and upscattering efficiency vary with latitude
because the current density and direction (along the magnetic field line)
depend on the position. The optical depth has a maximum and a minimum at the
equator and the poles (no current exists along the magnetic axis),
respectively. Thus, a hot spot presented at the pole would gain less energy
from particles in the magnetosphere, resulting in a soft spectrum. (ii) Photons
would be scattered away from the initial propagation directions and might move
into or out of the sight of view. The viewing effect should be taken into
account. (iii) The superposition of two thermal components at different
temperatures modifies the spectral profile further. To date, however, it is
impossible to incorporate all of these effects since it will involve too many
parameters. We emphasize that those extremely soft spectra at very low flux
levels cannot be fitted by the STEMS3D model nor the other normal Compton
scattering models (e.g., 1D RCS, 3D RCS, STEMS models). Nevertheless,
\cite{zane09} claimed that a softer spectrum is caused by a non-homogeneous
distribution of surface emission (effects (i) and (ii)). Meanwhile,
\cite{turolla11} also suggested that a surface thermal component dominated
spectrum can be reproduced by considering effects (i) and (iii).

\cite{beloborodov09} argued that the puzzling behavior of \xtej1810 can be
explained with a untwisting magnetosphere, which is divided into a current-free
(``cavity") region and a current-carrying bundle (``j-bundle") of field lines.
The hot spot in the polar region is interpreted as the footprint of the
j-bundle, which shrinks with time. As the magnetosphere untwists, the cavity
expands and the j-bundle becomes so narrow that only a small fraction of the
surface radiation can be scattered, producing very soft spectra in quiescence.
However, in this model it is difficult to explain that lack of flux enhancement
when glitches occurred in the stable sources. Alternatively, \cite{pons12}
interpreted transient behaviors as the magneto-thermal evolution, and suggested
that short X-ray bursts and glitches might always be accompanied by a flux
variation; nevertheless, the long-term flux of bright/stable magnetars (e.g.,
\axpu0142) cannot be enhanced significantly.

In some cases, magnetic configurations might be more complex than a dipole, and
multipoles fields are required to avoid too fast spin-down as we discussed
above. In addition, the theoretical calculation indicates that the output
luminosity from a globally twisted magnetosphere with $\Delta\phi > 1$ would be
at the order of $10^{36}$ erg s$^{-1}$, which is higher than the observed
\citep{thompson02, beloborodov09}. There are now some observational phenomena
in favor of partially, instead of globally, twisted magnetospheres
\citep[e.g.,][]{woods07, pavan09}, which have been considered in the literature
\citep{beloborodov09}. Finally, although we make an assumption of constant
velocity for charged particles, the particles should be decelerated via
scattering of photons, broadening their velocity distribution.

\acknowledgments{We thank the anonymous referee for helpful comments and
suggestions. We also thank Feryal \"Ozel for providing the highly magnetized NS
surface emission code, insightful discussions, and comments. S.S.W. is
supported by the Scientific and Technological Research Council of Turkey
(T\"{U}B\.{I}TAK) and EC-FP7 Marie Curie Actions-People-COFUND Brain
Circulation Scheme (2236).}

\end{document}